\documentclass[lettersize,journal]{IEEEtran}
\usepackage{amsmath,amsfonts}
\usepackage{algorithm}
\usepackage{algpseudocode}
\usepackage{array}
\usepackage{subcaption}
\usepackage{textcomp}
\usepackage{stfloats}
\usepackage{url}
\usepackage{verbatim}
\usepackage{graphicx}
\usepackage{cite}
\usepackage{soul}
\usepackage{xcolor}
\usepackage{tcolorbox}
\usepackage{afterpage}
\usepackage{threeparttable}
\usepackage{bm}

\newtcolorbox{highlighted}{colback=yellow!50,colframe=yellow!50,arc=0pt,outer arc=0pt}

\begin{document}

\title{Digital Twinning of a Pressurized Water Reactor Startup Operation and Partial Computational Offloading in In-network Computing-Assisted Multiaccess Edge Computing}

\author{Ibrahim Aliyu,~\IEEEmembership{Member,~IEEE}, Awwal M. Arigi, Tai-Won Um, and Jinsul Kim~\IEEEmembership{Member,~IEEE}
\thanks{This work was supported in part by the Institute of Information Communications Technology Planning Evaluation (IITP) grant funded by the Korean government (MSIT; RS-2024-00345030, Development of digital twin base network failure prevention and operation management automation technology, 50\%) and in part by the Innovative Human Resource Development for Local Intellectualization program through the IITP grant funded by the Korea government (MSIT; IITP-2024-RS-2022-00156287, 50\%).\textit{Corresponding authors: Tai-Won Um (email: stwum@jnu.ac.kr) and Jinsul Kim (email: jsworld@jnu.ac.kr)}.}
\thanks{Ibrahim Aliyu, and Jinsul Kim are with the Department of Intelligent Electronics and Computer Engineering, Chonnam National University, Gwangju 61186, Korea.}
\thanks{Awwal M. Arigi is with the Department of Humans and Automation, Institute for Energy Technology, 1777 Halden, Norway }
\thanks{Tai-Won Um is with the Graduate School of Data Science, Chonnam National University, Gwangju 61186, Korea.}}



\maketitle

\IEEEpubid{\begin{minipage}{\textwidth}\ \\[30pt] 
This work has been submitted to the IEEE for possible publication. Copyright may be transferred without notice, after which this version may no longer be accessible.
\end{minipage}}
\maketitle

\begin{abstract}
This paper addresses the challenge of representing complex human action (HA) in a nuclear power plant (NPP) digital twin (DT) and minimizing latency in partial computation offloading (PCO) in sixth-generation-enabled computing in the network (COIN) assisted multiaccess edge computing (MEC). Accurate HA representation in the DT-HA model is vital for modeling human interventions that are crucial for the safe and efficient operation of NPPs. In this context, DT-enabled COIN-assisted MEC harnesses DT (known as a cybertwin) capabilities to optimize resource allocation and reduce latency effectively. A two-stage approach is employed to address system complexity. First, a probabilistic graphical model (PGM) is introduced to capture HAs in the DT abstraction. In the PGM, HA and NPP asset-twin abstractions form coupled systems that evolve and interact through observable data and control input. Next, the underlying PCO problem is formulated as a multiuser game, where NPP assets can partially offload tasks to COIN and MEC. We propose a decentralized algorithm to optimize offloading decisions, offloading ratios, and resource allocation. The simulation results demonstrate the effectiveness of the proposed method in capturing complex HAs and optimal resource allocation in DT-enabled NPPs. 
\end{abstract}


\section{Introduction}
\IEEEPARstart{A}{dvances} in communication, artificial intelligence, and computing have been driving the digital transformation of nuclear power plants (NPPs). These developments have contributed to the emergence of digital twinning that replicates physical objects and their surroundings \cite{chua2023mobile}. Enabling digital twins (DTs) in NPPs enhances energy efficiency, enabling predictive maintenance, autonomous control, and safety analysis\cite{song2022online}.

In NPPs, human tasks/operations, such as pressurized water reactor (PWR) startup operations, include reactor operations, refueling, maintenance, shutdowns, and chemical control, which can be continuous or periodic. Digital twinning is essential for representing these processes accurately \cite{Arigi2024}. An accurate DT representation of human actions (HAs) is crucial for their impact on physical processes, safety measures, and critical event responses. The human-machine interface is also vital, covering interactions in control rooms, remote monitoring, diagnostic centers, and field devices.

Implementing a DT system in NPPs requires a diverse computing and communication infrastructure, from clusters to handheld devices \cite{yadav2021technical}. Sixth-generation (6G) ultra-reliable and low-latency computing (URLLC) capabilities are crucial for mission-critical DT applications \cite{huynh2022urllc}, and the computing in the network (also known as In-network computing, COIN) paradigm leverages network resources to minimize latency and enhance the quality of service \cite{aliyu2023dynamic}. Integrating a DT with these technologies offers opportunities and challenges for enhancing NPP reliability, availability, and maintainability at reduced costs.

\subsection{Related Work}
\subsubsection{Digital Twinning of HAs in NPPs} 
The DT is applied in various industries, including health, meteorology, education, manufacturing, transportation, energy, and business  \cite{rasheed2020}, \cite{rathore2021}. In the nuclear industry, DT adoption is limited due to high safety and security standards \cite{Arigi2024}. Nonetheless, research on DT adoption in nuclear systems has increased, focusing on technical challenges \cite{yadav2021}, \cite{kochunas2021}, \cite{prantikos2022}, safeguards and security  \cite{yadav2023}, applications for advanced reactors \cite{browning2022}, \cite{wilsdon2023}, and specific applications, such as flow-induced vibration prediction \cite{mohanty2021}. However, the consideration of HAs and procedures in DT remains largely unexplored.

\subsubsection{DT-enabled Task Offloading}
The COIN paradigm minimizes latency and improves the quality of experience using untapped network resources \cite{aliyutowards}, but may increase power consumption, creating a trade-off between delay and energy use. In this context, Partial subtask offloading is essential, especially when COIN coexists with edge computing solutions. Furthermore, digital twinning has been transforming resource allocation with intelligence, efficiency, and flexibility \cite{wu2021digital}. Recent studies have focused on DT-assisted task offloading in MEC \cite{do2022digital}, \cite{van2022edge}, \cite{li2022digital}, \cite{hao2023digital}. For instance, in \cite{liu2021digital}, the authors address DT-assisted offloading with mobile-edge server selection using the double deep Q network (DDQN), another study \cite{li2022digital}  focuses on energy optimization in MEC, and another \cite{hao2023digital} uses use of combinatorial optimization for computing overhead. These studies primarily address binary offloading, overlooking the need for partial offloading in the DT. Exploring partial offloading allows COIN and MEC to manage computationally intensive tasks collaboratively and efficiently.

\subsection{Motivation and Contributions}
The power startup scenario of an NPP involves increasing the reactor power (RP) to full capacity, requiring constant monitoring and complex manipulations due to disabled automatic and safety functions. Unlike standard full-power operations, this scenario demands more decision-making and poses a higher risk of human error due to changing system parameters and potential instability \cite{yadav2021technical}, \cite{Arigi2024}. Although artificial intelligence techniques have been studied for NPP control, these techniques have not addressed adapting the current systems for DT applications.

PCO is ideal for NPPs, as it efficiently distributes tasks across COIN nodes (CNs) and edge servers (ESs) , balancing latency and energy consumption through partial offloading \cite{kunze2022use}. This approach enhances overall system performance and operational safety by leveraging sensor data and factory monitoring to detect unusual HAs and support DT solutions.

Motivated by the aforementioned issues, this study focuses on the complex representation of human actions (HAs) within digital twins (DTs) while minimizing latency in partial computational offloading (PCO) within a 6G-powered COIN-assited MEC(C-MEC). Accurate representation of HAs in DTs is crucial for modeling human interventions that are vital for safely and efficiently operating NPPs. This work is the first to explore the digital twinning of HA in NPPs alongside the underlying DT-based network resource allocation problem in a collaborative C-MEC architecture. The primary contributions include the following:

\begin{itemize}
    \item We formulate the system utility maximization problem, jointly optimizing the digital twinning of HA-related PWR operations and resource allocation in the C-MEC, solving the problem in two stages.
    \item In the first stage, we apply a PGM to capture the intricacies of HAs in the DT abstraction, creating an interconnected system that evolves and interacts via observable data.
    \item In the second stage, we formulate the PCO problem as an exact potential game (EPG) and theoretically prove the existence of a Nash equilibrium (NE).
    \item Within the game, we formulate the offloading ratio and resource allocation (ORRA) problem as a Markov decision process (MDP) and employ the DDQN to predict and maximize future system utilities proactively.
    \item We conduct extensive experiments to evaluate the proposed digital twinning of the PWR operation and the underlying C-MEC-based network.
\end{itemize}

The rest of the paper is organized as follows. Section \ref{sec:system_model} presents the system model, including the DT-HA, DT-enabled C-MEC, communication, and computational models, and the problem formulation. Next, Section \ref{subsec: strateg} outlines the proposed approach involving the HAs and procedures using the PGM, and C-MEC resources for optimal communication and computing. Section \ref{sec:numerical} details the evaluation, and finally, Section \ref{sec: conclusion}  concludes the paper.
\begin{table}[ht]
    \centering
    \caption{Notation Summary}
    \label{table:notations}
    \scriptsize
    \begin{tabular}{p{1.5cm} p{5.5cm}}
        \hline
        \textbf{Notation} & \textbf{Description} \\
        \hline
        $I_m; C_m; T_m$ & Input size; CPU req.; Latency limit\\
        $\mathcal{M}; \mathcal{K}; R$ & Subsystems; COINs; Edge servers\\
        $\Phi_P; \lambda_{mk}; \aleph_m$ & Offloading ratio; Tasks at CNs; Tasks at ES\\
        $\Phi_L; \Phi_{\lambda}; \Phi_{\aleph}$ & Resource coeff.; CNs resource; ES resource\\
        $\omega_{m\Phi_L}(\bm{p}, \bm{n})$ & URLLC uplink rate\\
        $y_h$ & DT-HA state\\
        $U_m; \mathbf{s}; \phi(s)$ & Utility; Offloading profile; Potential function\\
        $\delta(\cdot)$ & DT-HA discrepancy\\
        $S_t; D_t; O_t$ & Physical assets; Digital state; Observed data\\
        $Q_t; R_t$ & Quantity of interest; Reward\\
        \hline
    \end{tabular}
\end{table}

\section{System model}
\label{sec:system_model}
This section presents the system model from the HAs and procedures and the COIN-assisted URLLC-based edge network. In addition, the problem emanating from the complex interaction of the two models is formalized. Fig. 1 illustrates the block diagram of the studied system, consisting of the DT-HA model and C-MEC services in which DT jointly optimised  NPP HAs and C-MEC resources. The main notations used in this paper are summarised in Table \ref{table:notations}

\subsection{Human Actions and Procedures in Digital Twin Nuclear Power Plants}
\label{subsec:DT-HA}
\subsubsection{Analysis of PWR startup operations}This study focuses on the startup operation of a typical Westinghouse three-loop PWR, following the general operating procedures (GOPs) \cite{Arigi2024}. The key GOPs include reactor coolant system filling and venting, cold shutdown, hot shutdown, hot standby to 2\% RP, and power operation and secondary system heat up and startup.

The critical part of the startup operation involves increasing the NPP power from 2\% to 100\% and transitioning the plant to normal conditions for electricity generation. Six major parameters serve as milestones: the pressurizer level (PL), reactor coolant temperature (RCT), reactor coolant pressure (RCP), steam generator (SG) pressure (SGP), SG level (SGL), and RP. These parameters are considered subsystems requiring communication and computing resources to operate.

Before reaching 2\% power, five GOPs are completed. After reaching 2\% power, only two GOPs are necessary: power operation greater than 2\% and secondary system heat up and startup. The former provides instructions for increasing the plant load from 2\% to 100\%, whereas the latter details steps for aligning and starting secondary systems. These procedures involve operating the rod controller, turbine load controller, feedwater (FW) pumps, condenser pumps, SG FW valves, and a synchronizer based on the planned rate of power increase.

In the first phase, control rods are withdrawn to 100\% (Banks A, B, and C to Step 228 and Bank D to Step 220) while increasing the boron concentration from 637 to 727 ppm to stabilize the power at 2\%. In the second phase, the boron concentration is reduced from 727 to 457 ppm, allowing the RP to rise to 100\%.

During the second phase, the turbine load controller maintains setpoints, including 1800 RPM at 10\% power, 2 MWe/min acceleration above 10\%, 100 MWe load at 10\% to 2\% power, and 200 MWe load at 20\% to 30\% power. In the first phase, FW Pump 1 operates, and FW Pumps 2 and 3 start at 40\% and 80\% power, respectively. Condenser Pumps 2 and 3 operate at 20\% and 50\% power, respectively. The synchronizer activates above 15\% power at 1800 RPM. The SG FW and pressurizer relief valves automatically control the levels. Various RCP, CVCS, makeup water, and nuclear service water systems support filling and venting the reactor coolant system.

\begin{figure}[!t]
  \centering
  \includegraphics[height=7cm,width=0.45\textwidth]{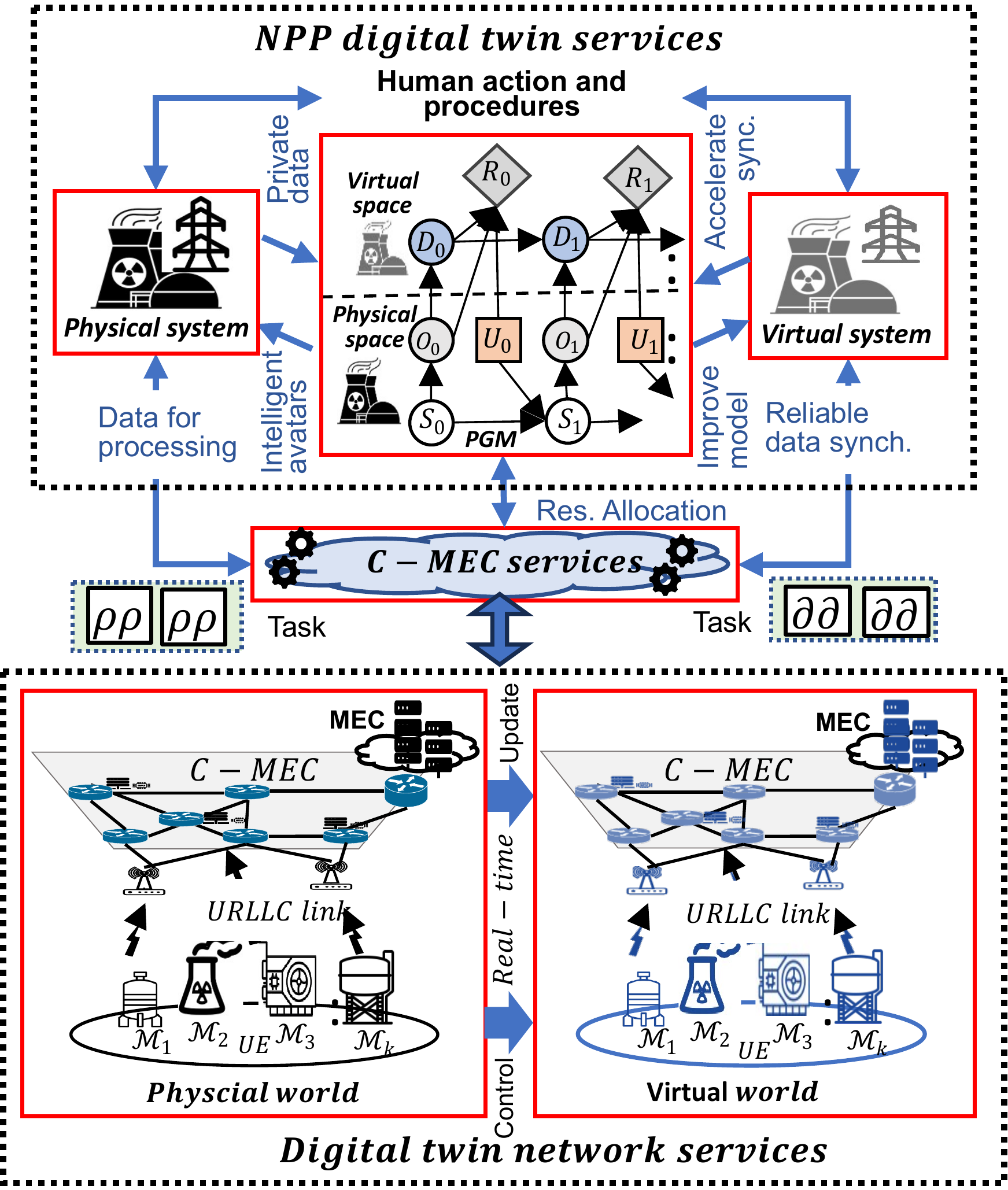}
  \caption{DT-enabled NPP and C-MEC system}
  \label{Block: PGM-EPG}
\end{figure}

\subsubsection{Operator Task DT Model}Accurately representing operator activities is crucial for ramping up power to 100\%. Human tasks are grouped into monitoring, decision-making, discrete control, and continuous control. The DT-HA model captures operator actions digitally, ensuring consistency and safety. An action profile lists the potential actions with likelihoods based on historical data. Environmental variables account for external conditions, highlighting interactions between the guidelines and such factors as temperature or noise.

\textit{Definition 1:} In the context of the proposed NPP model, the actions of each operator in the DT must precisely mirror real-world actions. For every operator action denoted as \( h \)  from a set of possible actions \( H \), the state of the DT, represented by \( y_h \), is governed by the following relation: 
 
\begin{equation}
\label{eq1}
y_h = f_h(S,A), \quad \forall h \in H
\end{equation}

where \( S \) and \( A \) represent the state/condition of the physical system and HA or decision-making in the system, respectively. Equation (1) ensures that the DT adheres to the operational guidelines, maintaining the highest standards of safety and operational accuracy.

The NPP startup operation depends on PL \( P_p \), RCT \( T_c \), RCP \( P_c \), SGP \( P_s \), SGL \( L_s \), and RP \( P_r \).

We let \( O_s (P_p, T_c, P_c, P_s, L_s, P_r) \) be an indicator function for operational constraints during the NPP startup, producing 1 when all operational constraints are met, and 0 otherwise. Similarly, \( S_s (P_p, T_c, P_c, P_s, L_s, P_r) \) is an indicator function for safety constraints during the NPP startup, returning 1 when all startup-specific safety constraints are satisfied and 0 otherwise.\\

\textit{Definition 2:} The general NPP operational constraint must be satisfied to ensure compliance with operational and safety regulations, preventing actions that could jeopardize NPP safety:
\begin{equation}
\label{eq2}
\begin{split}
O_s (P_p^O, T_c^O, P_c^O, P_s^O, L_s^O, P_r^O) \times \\
S_s (P_p^S, T_c^S, P_c^S, P_s^S, L_s^S, P_r^S) = 1
\end{split}
\end{equation}
The constraint ensures that all decisions align with the operational safety and procedural requirements of the NPP. This expression states that the operational and safety constraints must be satisfied for any given configuration of tasks, resources, and HAs. If either \( O_s \) or \( S_s \) is 0 (i.e., its respective constraints are not met), then the entire operation becomes 0, indicating that the configuration is invalid.

\subsection{COIN-assisted URLLC-based Edge Network Model Enabling NPP DT}
\label{subsec: C-MEC}
The C-MEC network architecture system model is illustrated in the lower part of Fig. \ref{Block: PGM-EPG}. The model consists of a physical layer comprising subsystems which represent the six major NPP subsystems and network resources, such as COIN-enabled CNs and the ES. This network infrastructure supports the operation of DT services by optimizing resource allocation, enabling the entire system via a real-time interaction mechanism.

We let $\mathcal{M} = \{1, 2, \ldots, M\}$ be a set of $M$ subsystems, $\mathcal{K} = \{1, 2, \ldots, K\}$ be a set of $K$ COIN CNs, and $R$ be the ES. The CNs and ES are associated with an access point (AP) to connect the subsystems. Further, URLLC short packet communication is employed between the subsystems and APs to ensure exceptionally reliable performance and low latency on the Internet of Things. The system model is as follows:

\subsubsection{Offloading Model in the C-MEC Network}
Regarding the time slot model, the subsystems and CNs are fixed within each time point and vary over different time slots. At each time slot $t$, each subsystem has a computational task characterized by $J_m = \{\eta_m, T_m^{\max}\}$, where $\eta_m = \frac{C_m}{I_m}$  represents the task complexity (cycle/bits), $I_m$ denotes the task size in bits, $C_m$ indicates the required CPU cycles (cycles) to execute the task, and $T_m^{\max}$ is the maximum tolerable latency for task $J_m$.

This scenario focuses on partial offloading to use parallel processing for latency reduction. We let $\Phi_P = \{\lambda_{mk}, \aleph_m\}$ be the offloading ratio variable, where $\lambda_{mk}$  denotes the portion executed at the CNs, and $\aleph_m = 1 - \lambda_m$  represents the portion of the task executed at the ES. Offloading resources are indicated by the variable $\Phi_L = \{\Phi_{\lambda}, \Phi_{\aleph}\}$ where $\Phi_{\lambda}$ and $\Phi_{\aleph}$ indicate the task execution resource locations at the CNs and ES, respectively. We assume the tasks are generated with high granularity, enabling partial offloading. For the task $J_m$, $I_m = \aleph_m I_m + \sum_{k\in K} \lambda_{mk} I_m$ and $C_m = \aleph_m C_m + \sum_{k\in K} \lambda_{mk} C_m$ satisfy $\aleph_m + \sum_{k\in K} \lambda_{mk} = 1$.

\subsubsection{C-MEC DT Model}
The DT services generate virtual replicas of physical systems, replicating the hardware, applications, and real-time data. The URLLC-based C-MEC DT is defined as
$\ DT = \{\tilde{\mathcal{M}},\ \tilde{\Phi}_L\},$ where $\{\tilde{\mathcal{M}},\ \tilde{\Phi}_L\}$ represents a virtual mirror of the system, including $M$ subsystems and $\tilde{\Phi}_L$ C-MEC computing resources (CNs and the ES). The DT layer, informed in real time, automates control via such services as data analysis, decision-making, and instant optimization, focusing on tasks that include offloading strategies and resource allocation.

The specific DT of each $m$-th subsystem (service module) is associated with a CN for processing, defined as $\ DT_m^{cn} = (f_m^{cn}, \tilde{f}_m^{cn}),$ where $f_m^{cn}$ denotes the estimated processing rate, and $\tilde{f}_m^{cn}$ quantifies the variation from the actual processing rate between the physical subsystem and its DT \cite{van2022urllc}. In the DT layer, the critical estimated processing rate, $f_m^{cn}$, mirrors subsystem behaviors, driving optimization decisions for device configurations. This rate is the focus of the optimization, with its deviation set as a predetermined percentage for simulations, following established practices \cite{do2022digital}.

Likewise, for the $\tilde{\Phi}_L$-th C-MEC computing resource (CNs and the ES), its	DT ($DT_{\tilde{\Phi}_L}^{cn}$) is formulated as $DT_{\tilde{\Phi}_L}^{cn} = (f_{\tilde{\Phi}_L}^{cm}, \tilde{f}_{\tilde{\Phi}_L}^{cm}),$ where $f_{\tilde{\Phi}_L}^{cm}$ signifies the estimated processing rate of the real C-MEC, and $\tilde{f}_{\tilde{\Phi}_L}^{cm}$ characterizes the disparity in the processing rate estimation compared to the actual C-MEC. The DT emulation of C-MEC (CNs and ES) provides valuable insight into C-MEC processing rates, facilitating the efficient allocation of computing resources and reducing processing latency through the offloading ratio and computing resource allocation adjustments.

\subsection{Communication Model}
\label{subsec: Comm}

The AP, with $L$ antennas serving $M$ single-antenna SMs, establishes channel connections with compute resource $\tilde{\Phi}_L$ represented by $\bm{h}_{m\tilde{\Phi}_L} = \sqrt{g_{m\tilde{\Phi}_L}} \cdot \bm{ \bar{h}}_{m\tilde{\Phi}_L}$, where $g_m$ is the large-scale channel coefficient and $\bm{ \bar{h}}_{m\tilde{\Phi}_L}$ is small-scale fading following $CN(0, \bm{I})$, where $CN(.,.)$ represents a complex circularly symmetric Gaussian distribution.A channel matrix $\bm{H}_{\tilde{\Phi}_L} = [\bm{h}_{1\tilde{\Phi}_L}, \bm{h}_{2\tilde{\Phi}_L}, \ldots, \bm{h}_{M\tilde{\Phi}_L}] \in \mathbb{C}^{L \times M}$ contains connections from $m$-th subsystems to the $\tilde{\Phi}_L$-th AP. Each subsystem has an allocated bandwidth, $b_m$. Match filtering and successive interference cancellation is employed to improve transmission performance \cite{fang2001performance}. 
Then, the signal-to-interference-plus-noise at the $\tilde{\Phi}_L$-th AP by the $m$-th subsystem is defined as $\gamma_{m\tilde{\Phi}_L}(p,n) = \frac{p_{m\tilde{\Phi}_L} \lVert h_{m\tilde{\Phi}_L} \rVert^2}{\mathcal{I}_{m\tilde{\Phi}_L}(p,n) + N_0},$
where $p_{m\tilde{\Phi}_L}$ denotes the transmission
power of the $m$-th subsystem, $N_0$ denotes noise power, $p = [p_{m\tilde{\Phi}_L}]_{m=1}^M$, and $\mathcal{I}_{m\tilde{\Phi}_L}(p,n) = \sum_{n>m}^{M} p_{n\Phi_L} \frac{\lvert h_{m\Phi_L}^H h_{n\Phi_L} \rvert^2}{\lVert h_{m\Phi_L} \rVert^2} $ represents the interference imposed by subsystems $n > m$. Thus, the uplink URLLC transmission rate is expressed as follows \cite{ren2020joint,she2017radio}:
\begin{equation}
\begin{split}
    \omega_{m\Phi_L}(\bm{p}, \bm{n}) &\approx B \log_2\left[1 + \gamma_{m\Phi_L}(\bm{p}, \bm{n})\right] \\
    &\quad - B\sqrt{\frac{V_{m\Phi_L}(\bm{p}, \bm{n})}{N}} \frac{Q^{-1}(\epsilon)}{\ln 2}, \label{eq:uplink_rate}
\end{split}
\end{equation}
where $B$ represents the system bandwidth, $\epsilon$ characterizes the likelihood of decoding errors, $\gamma_{m\tilde{\Phi}_L}(p,n)$ stands for the Signal-to-Noise Ratio (SNR) observed by the $m$-th subsystem, $Q^{-1}(.)$ denotes the reverse function of $Q(x) = \frac{1}{\sqrt{2\pi}} \int_x^{\infty} e^{-t^2/2} \,dt$, and $V_{m\tilde{\Phi}_L}$ is the channel dispersion given as $V_{m\tilde{\Phi}_L}(p,n) = 1 - \left[1 + \gamma_{m\tilde{\Phi}_L}(\bm{p}, \bm{n})\right]^{-2}$. 
The uplink wireless transmission latency from the $m$-th subsystem to the $\tilde{\Phi}_L$-th C-MEC resource can be expressed as follows:
\begin{equation}
    T_{m\tilde{\Phi}_L}^{\text{CO}}(\bm{p}, \bm{n},\tilde{\Phi}_L) = \max_{\forall\Phi_L} \left\{\frac{\Phi_L I_m}{\aleph_{m\Phi_L}(\bm{p}, \bm{n})}\right\}. \label{eq:uplink_latency}
\end{equation}

\subsection{Computational Model}
\label{subsec: Compt.}
In the computational model, each subsystem generates a granular computation task $J_m$ in which a portion can be executed by the CNs and another portion at the ES. The model is defined as follows:

\subsubsection{COIN Node Processing}
For the COIN node, the task $J_m$ portion $\lambda_{mk}$ is executed by CNs with the estimated processing rate $f_m^{cn}$. Thus, the estimated CN execution latency is as follows:
\begin{equation}
    \tilde{T}_{mk}^{cn}(\lambda_{mk},f_m^{cn}) = \max_{\forall k \in K} \left\{\frac{\lambda_{mk} C_m}{f_m^{cn}}\right\}. \label{eq:coin_processing_time}
\end{equation}

If the discrepancy between the actual $k$th CN and its DT can be predetermined, the gap in computing latency between real-world performance and DT predictions can be estimated as follows:
\begin{equation}
    \Delta T_{mk}^{cn}(\lambda_{mk},f_m^{cn}) = \frac{\lambda_{mk} C_m \tilde{f}_{mk}^{cn}}{f_m^{cn} (f_m^{cn} - \tilde{f}_{mk}^{cn})}. \label{eq:delta_latency}
\end{equation}

Thus, the actual CN processing time is $ T_{mk}^{cn} = \Delta T_{mk}^{cn} + \tilde{T}_{mk}^{cn}$. The total latency, including the transmission and computing latency is 
\begin{equation}
T_m^{cnT} = T_{mk}^{cn} + T_{m\Phi_L}^{CO}
\end{equation}

\subsubsection{MEC Processing}
The task $J_m$ portion $\aleph_m$ executed by the ES with the estimated processing rate of $f_m^{em}$ incurs the following latency:
\begin{equation}
\tilde{T}_m^{em}(\aleph_m,f_m^{em}) = \frac{\aleph_m C_m}{f_m^{em}}. \label{eq:mec_processing_time}
\end{equation}

The latency gap $\Delta T_m^{em}$ between the real latency and the DT is estimated as follows:
\begin{equation}
\Delta T_m^{em}(\aleph_m,f_m^{em}) = \frac{\aleph_m C_m \tilde{f}_m^{em}}{f_m^{em}(\tilde{f}_m^{em}-f_m^{em})} \quad
\end{equation}

Thus, the actual latency for task execution at $
T_m^{em} = \Delta T_m^{em} + \tilde{T}_m^{em} \quad $. The total delay at MEC is
\begin{equation}
T_m^{emT} = T_{m\Phi_L}^{CO} + T_m^{em}
\end{equation}

\subsubsection{ Latency Model}
The total end-to-end (e2e) DT latency within the system includes the  CN processing latency, task offloading transmission latency, and ES processing latency. Thus, the e2e DT latency is expressed as
\( T_m^{e2e} = T_m^{kcn} + T_{m\Phi_L}^{CO} + T_m^{em} = \max_{\forall k\in K}\left\{\frac{\lambda_{mk} C_m}{f_m^{kcn}-\tilde{f}_m^{kcn}}\right\} + \max_{\forall\Phi_L}\left\{\frac{\Phi_L I_m}{\omega_{m\Phi_L}(p,n)}\right\} + \frac{\aleph_m C_m}{f_m^{em}-\tilde{f}_m^{em}} \).

\subsection{Problem Formulation}
\label{subsec: Prob.}
We let \( S_m=\{s_{m0},s_{m1},s_{m2}, \ldots, s_{mK} \mid s_{mj} \in \{0,1\}\} \) denote the offloading strategies for subsystem \( m \). The offloading strategy profile of all subsystems is denoted as \( \mathbf{s}=\{s_m \mid s_m \in S_m, m \in M\} \), where \( s_m=s_{mj}=1 \) suggests that subsystem \( m \) accomplishes its task via decision \( j \), otherwise \( s_m=s_{m0}=0 \). \( s_{m0} \) indicates the decision variable for task execution at the ES while \( s_{mk} \) are executed at the CN \( k \). From the subsystem perspective, we define the subsystem \( m \) utility as the difference between the reduced latency due to offloading and the computational cost as follows:
\begin{equation}
\label{qtn 11}
 U_m=\sum_{j \in K \cup \{0\}} s_{mj} [g_t (T_m^{em} - T_m^{e2e}) - p_j \Phi_j C_m]   
\end{equation}
where \( g_t \) is the unit gain latency reduction and \( p_j \) is proportional to computing capacity, indicating offloading cost per workload at node \( j \).

The primary objective is to maximize the system utility by minimizing the overall system latency, considering HAs and task offloading. The problem formulation is as follows:

\begin{align}
    \mathcal P: & \max_{s, \Phi, \beta, H} \sum_{m \in M} U_m  \\
    \nonumber \\
    \text{s.t.} & \sum_{j \in K \cup \{0\}} s_{mj} \leq 1 \quad \label{12a} \tag{12a} \\
    & \sum_{m \in M} s_{mj} \leq 1 \quad \label{12b} \tag{12b} \\
    & s_{mj} T_m^{e2e} \leq T_{\text{max}} \quad \label{12c} \tag{12c} \\
    & \sum_{m \in M} \beta_m \leq 1 \quad \label{12d} \tag{12d} \\
    & y_h = f_h (S,A), \forall h \in H \quad \label{12e} \tag{12e} \\
    & O_s (P_p, T_c, P_c, P_s, L_s, P_r) \times \nonumber\\ 
    & S_s (P_p, T_c, P_c, P_s, L_s, P_r) = 1 \quad \label{12f} \tag{12f} \\
    & s_{mj} \in \{0,1\}, 0 \leq \Phi, \beta \leq 1, \forall h \in H, y_h \in D,\nonumber\\
    & \forall m \in M, j \in K \cup \{0\} \quad \label{12g}\tag{12g}
\end{align}

Constraint (\ref{12a}) ensures each task is partially offloaded to at most one  node. Constraint (\ref{12b}) manages the associations between the subsystem and COIN node. Constraint (\ref{12c})  enforce latency requirements and (\ref{12d}) ensures resource allocation is within CN capacity. Constraint (\ref{12e}) ensures the DT state matches the current system and HAs and (\ref{12f}) aligns decisions with operational and safety constraints. Finally, (\ref{12g}) denotes optimization constraints.

The objective function is non-convex due to binary decisions, nonlinear relationships, and multiplicative interactions. The function intractable to solve directly because it involves interactive DT-HA and PCO in a C-MEC cybertwin over multiple time slots and lacks subsystem request transition probabilities.

\section{Proposed Strategy}
\label{subsec: strateg}
Due to the intractability of problem $\mathcal{P}$, the DT problem is decomposed into three subproblems: offloading ratio optimization, resource allocation, and HA digital twinning integration. We propose a decentralized game-theoretical approach to represent the DT-HA model and minimize latency in PCO within the C-MEC environment. The PGM captures the intricacies of the DT-HA model, where the DT state reflects the system state and determines HAs, evolving through observable data and control inputs.

Next, the offloading decision is modeled as a strategic game with HA constraints to determine subsystem utility. As a rational player, each subsystem devises its offloading strategy based on others' strategies. The DDQN refines the ORRA, optimizing PCO decisions, offloading ratios, and resource allocation under a given DT-HA state. Fig. 2 illustrates the proposed scenario, divided into two parts: the DT-HA using the PGM and the C-MEC services, which provide optimal communication and computing resources for efficient DT operation. The operational flow involves four steps: initializing system state information, estimating the digital twin (DT) of the cyber twin and PWR human actions, maximizing user utility via the GT-PCO module, and updating the system with optimal task offloading and resource allocation. This approach enhances automation, accuracy, safety, and efficiency in NPP operations. 

\begin{figure}[!t]
  \centering
  \includegraphics[width=0.4\textwidth]{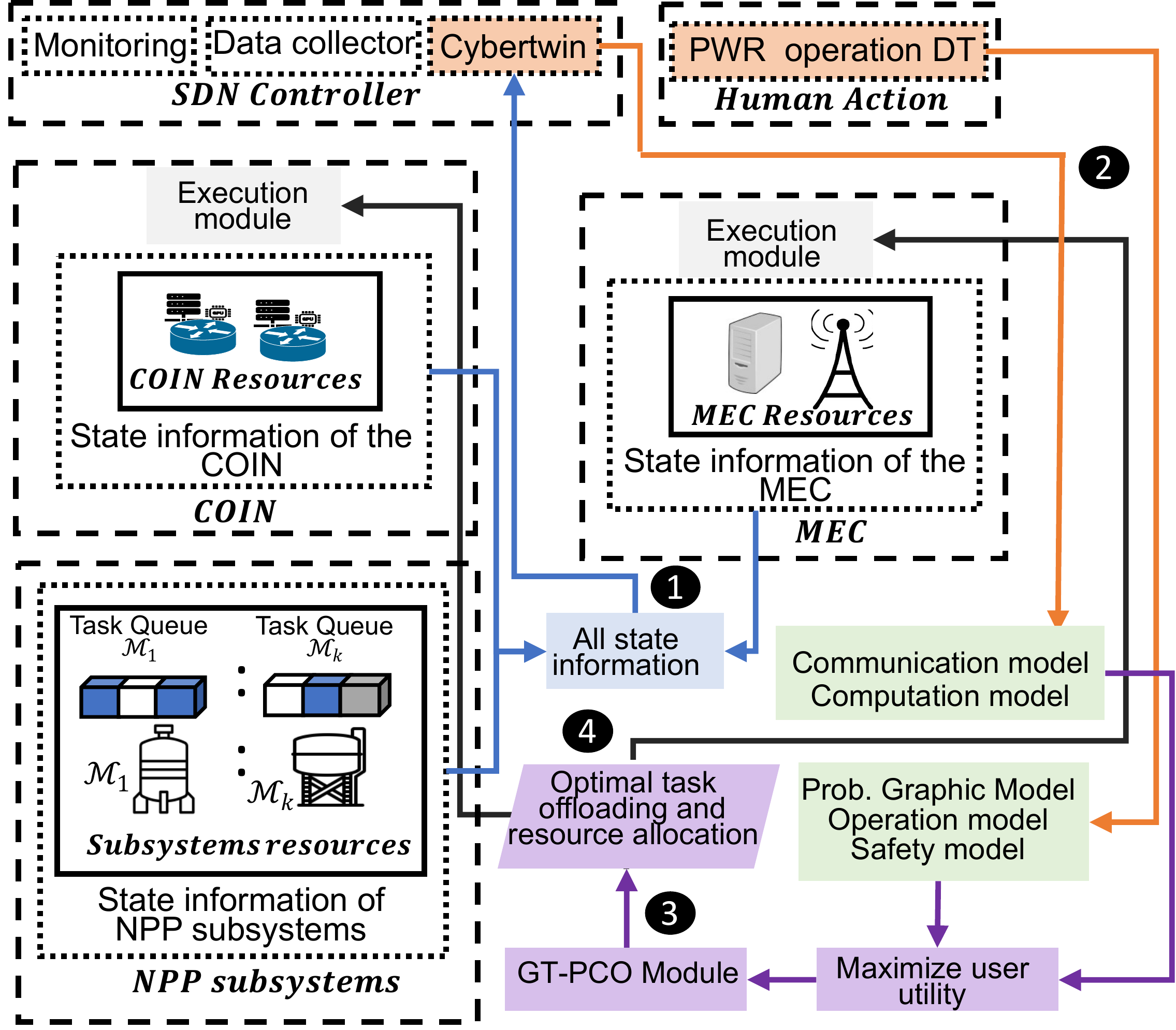}
  \caption{Proposed digital twinning of the nuclear power plant (NPP) human actions and in-network computing (COIN)-assisted multiaccess edge computing (MEC) (C-MEC)}
  \label{PGM-EPG}
\end{figure}

\subsection{PGM for Digital Twining of Human Actions Procedures}
\label{subsec: DT-HA PGM}
The DT-HA model aims to align the DT and HAs/system states accurately. The DT-HA model mirrors operator actions, adhering to standards and historical data for predictive insight. System feedback tracks plant responses, and environmental variables consider external factors. Problem \( \mathcal{P} \) is reformulated as follows to address the DT-HA problem: 

\begin{align}
    & \mathcal{P}_1: \min_{h \in H} \delta(y_h, f_h) \nonumber \\
    & \text{s.t.} \ \text{(12f)}, \ \forall h \in H, \ y_h \in D.
\end{align}

\noindent where \( \delta(y_h, f_h) \) represents the discrepancy function that quantifies the difference between the current state \( y_h \) of the DT and the state predicted by the HAs \( f_h \). The \( \mathcal{P}_1 \) problem aims to estimate the expected state due to HAs while minimizing the discrepancy with the actual DT state, adhering to the operational and safety guidelines. Solving \( \mathcal{P}_1 \) identifies the optimal HAs and system configurations that enhance system performance while maintaining the accuracy of the DT.

Next, \( \mathcal{P}_1 \) is translated to a PGM, where the DT of the NPP uses this model to generate experimental data for calibration and performance evaluation. The calibrated DT operates alongside the physical asset, assimilating the sensed data to update its internal models.

The PGM for the DT, proposed by Kapteyn et al. \cite{kapteyn2021}, considers the physical asset and its DT, evolving through time. The DT estimates the current and future states of the physical asset based on observational data, providing optimal control inputs to direct the physical asset to the desired states.

The proposed model uses six variables: the physical state \( S_t \), observable data \( O_t \), digital state \( D_t \), control inputs \( U_t \), quantities of interest \( Q_t \), and reward \( R_t \). These variables represent the state of physical assets, parameters defining DT models, available information on the physical asset state, actions influencing the digital asset, estimated parameters via model outputs, and the overall performance of the asset-twin system, respectively.

The PGM represents the asset-twin system structure by encoding the interaction and evolution of these quantities. The PGM covers the data-to-decision flow from sensing to action. The conditional independence structure of the model allows the factorization of the joint distributions over model variables.

The system is modeled using a dynamic Bayesian network with decision nodes, representing the system from the initial time step \( t=0 \) to the current time step \( t=t_c \) and future time step \( t=t_p \). The graph nodes are random variables denoting each quantity at discrete time points, with uppercase letters representing variables and lowercase letters denoting their values \( D_t \sim p(d_t) \). The graph edges encode dependencies between variables via conditional probabilities or deterministic functions.

\begin{figure}
\centering
\includegraphics[width=0.8\columnwidth]{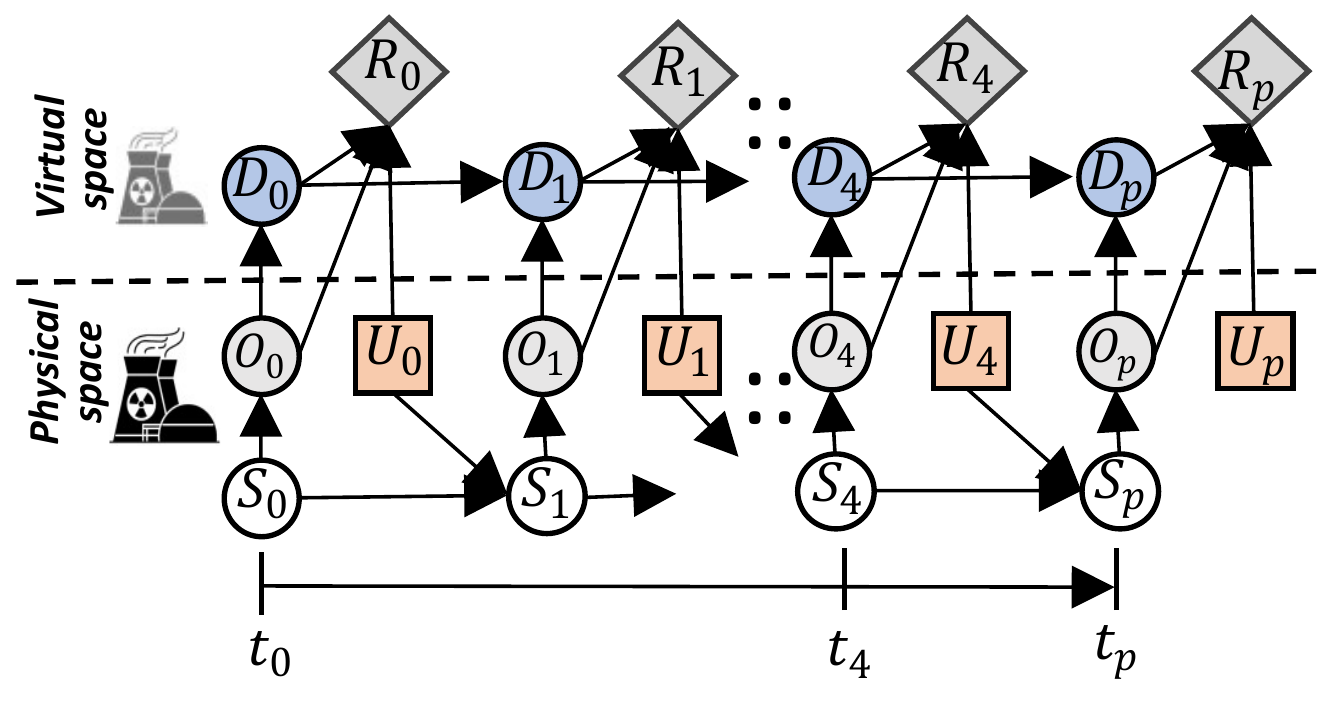}
\caption{Dynamic decision network for the physical asset and digital twin of the nuclear power plant. Nodes with bold outlines are observed quantities; others are estimated using a probability distribution. Edges illustrate conditional dependencies}
\label{block:PGM_evol}
\end{figure}

Fig. \ref{block:PGM_evol} illustrates the system evolution. The time evolution of the physical asset states \( S_t \sim p(s_t) \)  is defined as \( \{ S_t \} = \{ S_0, S_1, S_2, S_3, S_4 \} \), where \( S_0, S_1, S_2, S_3, S_4 \) represent the reactor coolant system filling and venting, cold shutdown, hot shutdown, hot standby, and power generation, respectively. The corresponding digital state \( D_t \sim p(d_t) \) is given as \( \{ D_t \} = \{ D_0, D_1, D_2, D_3, D_4 \} \). The interaction between the physical asset and its DT is facilitated through the information flow. The information flow in the form of observational data \( o_t \), from physical assets to its DT, updates the corresponding digital state in the process. The quantities of interest \( Q_t \sim p(q_t) \) are computed by the updated DT model. For our scenario, the \( Q_t = O_t \), reflecting interest in all observable data for system operation. The digital state and computed quantity of interest control the input \( u_t \) from the DT to the physical asset. The reward for the time step \( R_t \sim p(r_t) \) is determined by all these quantities.
The graphical approach enables defining known or assumed conditional independence. The model encodes the physical and digital state based on the Markov assumption observable through the data. By exploring the conditional independence, the joint distributions over variables can be factorized in the model as follows:

\begin{align}
    p(o_0, \ldots, o_{t_c}, \ldots, u_0, \ldots, u_{t_c}) &= \prod_{t=0}^{t_c} \left[\phi_t^{\text{update}} \phi_t^{\text{QoI}} \phi_t^{\text{evaluation}}\right], \label{eq:main} 
\end{align}
\noindent where
\begin{align}
    \phi_t^{\text{update}} &= p(D_{t-1}, U_{t-1} = u_{t-1}, O_{t-1} = o_t), \label{eq:update} \\
    \phi_t^{\text{QoI}} &= p(D_t), \label{eq:qoi} \\
    \phi_t^{\text{evaluation}} &= p(D_t, Q_t, U_t = u_t, O_t = o_t). \label{eq:evaluation}
\end{align}
The state can be predicted for \( t_p \) by extending the state to include the digital state, quantity of interest, and reward variable.
The factors in \ref{eq:main} are conditional probability distributions characterizing interactions in the DTs, as suggested by \cite{kapteyn2021}. The Bayesian inference algorithm uses this equation for asset monitoring, prediction, and optimization. A PGM is instantiated for the DT of a PWR NPP to optimize human operations and ensure safety. The PGM calibrates, operates, and updates models to reflect the current state of the NPP.

The NPP DT life cycle includes two stages: calibration and operation. The first stage calibrates the DT using the PGM for scalability and repeatability. The second stage involves deploying the calibrated twin to monitor, predict, and optimize NPP operations, maintaining operational parameters. The physical state, \( S \), of the NPP assets corresponds to the stages of GOPs during startup operation. These procedures and their states are captured by creating, calibrating, and evolving the DT

\subsection{Multisubsystem Computation Offloading Game}
\label{subsec: C-MEC game}
The multiuser computation offloading game can be defined as $G=\{M, (S_m)_{m\in \mathcal{M}}, (U_m)_{m\in \mathcal{M}}$, where $S_m$ denotes the set of offloading strategies for subsystem $m$, and $U_m(s_m,\bm{s}_{(-m)})$ represents the utility function, considering the set of offloading strategies. In addition, $s_{-m}=(s_1,\ldots,s_{(m-1)},s_{(m+1)},\ldots,s_M)$ represents the offloading strategies of all subsystem except the $m$th, where each elects the most advantageous strategy that enhances its individual utility. The game is considered to achieve a state of NE when no subsystem can further improve its utility by altering its offloading choice. \\
\textit{Definition 1:} A strategy $s^*=(s_1^*,s_2^*,\ldots,s_M^*)$ is the NE of the game $G$ if it adheres to

\begin{equation}
\begin{gathered}
U_m(s_m^*,\bm{s}_{(-m)}^*) \geq U_m(s_m,\bm{s}_{(-m)}^*), \\
\forall m\in M, \forall s_m\in S_m.
\end{gathered}
\end{equation}

Based on \cite{pham2022partial}, the game $G$ is an EPG by formulating the potential function as follows:

\begin{equation}
\begin{aligned}
\phi(s) = &s_{m0} \sum_{m\in M}\left[ R_{m0} + (1-s_{m0})\left(\sum_{j\in K}s_{mj}R_{mj} \right. \right.\\
&\left. \left. + \sum_{m'\neq m} R_{m'0}\right)\right] \quad 
\end{aligned}
\end{equation}
where $R_{mj} = g_t(T_m^{em} - T_m^{kcn}) - p_j \Phi_j C_m$.
For ease of proof, the expression $\phi(s_m,s_{(-m)})$ is given as follows:
\begin{equation}
\begin{aligned}
\phi(s_m,s_{(-m)}) = \, 
s_{m0} \sum_{m\in M}\bigg[ R_{m0} + (1-s_{m0}) \\ 
\bigg(\sum_{j\in K}s_{mj}R_{mj}
 + \sum_{m'\neq m} R_{m'0}\bigg)\bigg]
\end{aligned}
\end{equation}

\textit{Remark 1:} Game $G$ with the potential function $\phi(s)$ is an EPG and capable of reaching an NE in a finite number of iterations.

\textit{Proof:}  See Appendix \ref{appendix: NE_prove}

\subsection{Deep Double Q Network for Optimal Offloading Ratio and Resource Allocation}
The joint optimization of the ORRA problem can be reformulated to maximize the utility as follows:

\begin{align}
    & \mathcal{P}_2: & & \min_{\Phi,\beta} \sum_{m\in M} p_j \Phi_j C_m - (T_m^{em} - T_m^{e2e}) \nonumber \\
    & &\text{s.t.} &\ \text{(12c), (12d)}, \  0 \leq \Phi, \beta \leq 1, \quad \forall m\in M. \label{eq:constraint_24c}
\end{align}

For any time slot $(t+1)$ given the user offloading request $\mu_{(t+1)}$, the optimal offloading ratio $\Phi_{(t+1)}$ and resource allocation $\beta_{(t+1)}$ can be solved. However, $\mu_{(t+1)}$ is unknown due to the unknown user request transition probabilities. The DDQN is employed to capture the user request model and predict the optimal task offloading ratio and corresponding resource allocation of time slot $(t+1)$ based on the system state at slot $t$.

We formulate the $\mathcal{P}_2$ as a Markov Decision Process (MDP) and elaborate on the state, action, and reward as follows:\\
\textbf{State:} The user request state at time slot $t$ is denoted by $S_t = \mu_t \in (F+1)^M$, where $F$ is the number of tasks. 

\noindent\textbf{Action:} The action at time slot $t$ is the ORRA $A_t = \Phi_{t+1}, \beta_{t+1} \in [0,1]^M$.

\noindent\textbf{Reward:} The reward at time $t$ is defined as the utility savings in time $(t+1)$, denoted by $R_{t+1}$. This savings is calculated as the difference between the utility derived from the optimal partial ORRA and that from the full ORRA simultaneously.

\subsection{Game-Theoretic Offloading Framework}
The game-theoretic offloading framework (Algorithm 1) solves the PCO decision problem ($\mathcal{P}$) using the future optimal ORRA problem $(\mathcal{P}_2)$ for efficient computation offloading under DT-HA constraint in $(\mathcal{P}_1)$. The base station (BS) acts as the central hub in its operation, assimilating real-time data, such as connection statuses and subsystem strategies. Initially, service modules (subsystems) lean toward MEC offloading. However, as the iterations progress, the subsystem refines its offloading strategy based on feedback from the BS. This iterative exchange continues until the subsystem seeks no further updates, indicating an NE. The computational complexity of the game-theoretic offloading framework is represented as $O(C1 \times N )$, where $C1$ is the iteration count for the DDQN.

\begin{algorithm}[!t]
    \caption{Multiuser PCO}
    \label{alg:multiuser_pco}   
    \scriptsize
    \begin{algorithmic}[1]
        \State \textbf{Initialization:} Subsystem $k \in K$ initializes its PCO decision toward offloading to the MEC server.
        \For{decision slot $t$:}
            \For{subsystem $k \in K$:}
                \State Obtain real-time system environment from the BS.
                \If{if available DTs exist, then}
                    \State Determine $U_{ij}$ according to equations pertaining to the ORRA Problem ($\mathcal{P}_1$).
                \EndIf
                \State Obtain the best optimal strategy such that $U_m(s_m',s_{-m})=\arg\min_{S_{(k,t)}} U_{mj}$
                \If{$U_m(s_m',s_{-m}) > U_m(s_m,s_{-m})$}
                    \State The updated PCO strategy, $S_{(k,t)} = S_{(k,t)}^*$, is sent to the BS and stored in $M(t)$.
                \Else
                    \State Retain the previous strategy $s_m$.
                \EndIf
            \EndFor
            \If{$M(t) \neq \emptyset$}
                \State The subsystems in $M(t)$ vie for the next update opportunity.
                \If{subsystem $i$ wins}
                    \State Broadcast the update to all subsystems: $s_m(t) = s_m'$.
                \Else
                    \State Retain the previous strategy $s_m$.
                \EndIf
            \EndIf
        \EndFor
        \Repeat
            \Until an END message is received
        \State \textbf{Return:} The optimal offloading strategies $s^*$.
    \end{algorithmic}
\end{algorithm}

\section{Numerical Results}
\label{sec:numerical}

\subsection{Data-driven Calibration and Evolution of a Digital Twin Human Action Model}
The data-driven approach uses a state-transition infinite state automation for DT capabilities, including monitoring, prediction, and optimization, ensuring continuous model improvement. A PGM was employed to represent HAs in an NPP. The model calibration and updates reflect the state of the NPP based on HAs. The DT life cycle includes calibration and operational phases, enabling predictive simulation and autonomous operation according to the guidelines.
In NPP operations, the physical state \(S_t\) evolves due to HAs, whereas the digital state \(D_t\) models these actions, accounting for subsystem differences. The critical system variables, such as the PL, RCT, RCP, SGP, SGL, and RP, are defined across various operational states, as listed in Table I. Intermediate values are computed using linear interpolation:

\begin{equation}
\mathbf{x}_s = \mathbf{x}_i + \frac{s}{n} (\mathbf{x}_{i+1} - \mathbf{x}_i)
\end{equation}
where \(\mathbf{x}_i\) and \(\mathbf{x}_{i+1}\) denote values at states \(S_i\) and \(S_{i+1}\), respectively, and \(s\) is the step out of \(n\) steps.

This interpolation creates a high-resolution dataset for smooth state transitions. State labels are adjusted based on RP values, enhancing the precision of the NPP DT in monitoring and controlling plant operations. Table \ref{table:system-variables} presents the control input mapping and estimated observable data.

\begin{table}[ht]
    \centering
    \caption{System variables}
    \label{table:system-variables}
    \begin{threeparttable}
        \begin{tabular}{|c|c|c|c|c|c|c|c|}
        \hline
        \textbf{S\textsubscript{t}} & \textbf{D\textsubscript{t}} & \textbf{PL} & \textbf{RCT} & \textbf{RCP} & \textbf{SGP} & \textbf{SGL} & \textbf{RP} \\
        \hline
        S\textsubscript{0} & D\textsubscript{0} & 100 & 60 & 27 & 1 & 100 & 0 \\
        S\textsubscript{1} & D\textsubscript{1} & 100 & 176 & 27 & 1 & 60 & 0 \\
        S\textsubscript{2} & D\textsubscript{2} & 20 & 176 & 156 & 76.6 & 100 & 0 \\
        S\textsubscript{3} & D\textsubscript{3} & 50 & 294 & 157 & 76.6 & 50 & 2 \\
        S\textsubscript{4} & D\textsubscript{4} & 50 & 308 & 157 & 76.6 & 50 & 100 \\
        \hline
        \end{tabular}
        \begin{tablenotes}
            \item Notes: PL: pressurizer level, RCT: reactor coolant temperature, RCP: reactor coolant pressure, SGP: steam generator pressure, SGL: steam generator level, RP: reactor power
        \end{tablenotes}
    \end{threeparttable}
\end{table}

\begin{table}[ht]
    \centering
    \caption{Control input mapping}
    \label{table:control-input}
    \begin{tabular}{|c|cccc|c|ccc|ccc|}
    \hline
    \textbf{U\textsubscript{t}} & \multicolumn{4}{c|}{\textbf{Rod Ctrl}} & \textbf{Boron Ctrl} & \multicolumn{3}{c|}{\textbf{Water feed}} & \multicolumn{3}{c|}{\textbf{Cond. pump}} \\
    \cline{2-5} \cline{7-12}
    & \textbf{A} & \textbf{B} & \textbf{C} & \textbf{D} & & \textbf{1} & \textbf{2} & \textbf{3} & \textbf{1} & \textbf{2} & \textbf{3} \\
    \hline
    U\textsubscript{0} & 1 & 1 & 0 & 0 & Increase & 1 & 0 & 0 & 1 & 0 & 0 \\
    U\textsubscript{1} & 1 & 1 & 0 & 0 & Increase & 1 & 0 & 0 & 1 & 0 & 0 \\
    U\textsubscript{2} & 0 & 1 & 1 & 0 & Increase & 1 & 0 & 0 & 1 & 0 & 0 \\
    U\textsubscript{3} & 0 & 0 & 1 & 1 & Increase & 1 & 0 & 0 & 1 & 0 & 0 \\
    U\textsubscript{4} & 1 & 1 & 1 & 1 & Decrease & 1 & 1 & 1 & 1 & 1 & 1 \\
    \hline
    \end{tabular}
\end{table}

\textbf{Calibration phase}. In the calibration phase, the DT is customized to reflect subsystem responses during energy production. This approach relies on the operational guidelines and historical data, capturing subsystem parameters and uncertainties.

At each time step \( t \), control actions \( U_t = u_t \) involve controllers, such as the rod controller and turbine controller, to maintain system stability. Parameters are calibrated stage-by-stage based on the historical data and GOP dependencies. The proposed model targets the observable data \( P(Q_t \mid D_t, O_t = o_t) \)  and evaluates the estimated reward \( P(R_t) \) at each stage.

At step \( t=0 \) (system cooling and venting), the RP is set to 0\%. Control actions \( U_t \) include rod control (Banks A and B) withdrawal to 65\% and 30\%, corresponding to \( U_1 \). The posterior \( P(D_0 \mid O_0, U_t=u_0) \) dynamically sets the subsystem states. Step \( t=1 \) calibrates the cold shutdown, with the RP still at 0\%, RCT at 176, and SGL at 60. The posterior estimate is \( P(D_1 \mid D_0, O_1, U_t=u_1) \). In step \( t=2 \) (hot shutdown), the PL is 20, the RCT is 176, and the RCP, SGL, and SGP increase to 6, 76.6, and 100, respectively. The posterior is updated to\( P(D_2 \mid D_1, O_2, U_t=u_2) \). Steps \( t=3 \) and \( t=4 \) calibrate the hot standby and power generation stages. Here, the PL, RCT, and SGL were maintained at 50, 157, 76.6, and 100, respectively. The final stage signifies the energy production and plant operation beyond 2\%, focusing on the power increase proportional to the energy production.

\textbf{Operation Phase}.
The calibrated DT is deployed with the NPP during the PWR startup following strict guidelines. This phase demonstrates how the PGM manages uncertainties from calibration and assimilation and extends the capabilities of the DT for planning, predicting, and evaluating human operations.
At time steps \( t=4 \) and $5$, during power production, the model uses real NPP data to predict the control inputs, observable data, state, and reward accurately, enhancing system confidence. The DT determines the expected system state and control operations based on operational guidelines. Optimal control inputs are identified through the assimilation and evolution capabilities of the model. The DT dynamically estimates the operational state of the NPP by assimilating the observable data and adjusting the predictions, with the data for each time step described as subsystem readings. For each subsystem \( m \), the observable data \( O_t \) at time  \( t \) is given by the following:

\begin{equation}
O_t = Q_t = \{O_t^m\}_{m=1}^M
\end{equation}

where \( M=6 \) for simplicity.

The digital state update factor \( \phi_t^{\text{update}} \) signifies the digital state update at each time step given the previous digital state and control input at time \( t-1 \), and the current observable input data:

\begin{equation}
\phi_t^{\text{update}} \propto \phi_t^{\text{dynamics}} \phi_t^{\text{assimilation}}
\end{equation}

where 
\begin{equation}
\phi_t^{\text{dynamics}} = P(S_{t} \mid  S_{t-1}, U_{t-1} = u_t)
\end{equation}
\begin{equation}
\phi_t^{\text{assimilation}} = P(O_t=o_t,S_t)
\end{equation}

The DT update applies the Bayes rule to update the probability distribution \( P(S_t \mid O_{1:t}) \) given the new observation \( O_t \):

\begin{equation}
P(S_t \mid O_{1:t}) = \frac{P(O_t \mid S_t) P(S_t \mid O_{1:t-1})}{P(O_t \mid O_{1:t-1})}
\end{equation}

The model updates \( P(O_t=o_t, S_t) \) using new observations, calculating the likelihood of observed data given the current state and updating the state belief accordingly.

The assimilation factor \( \phi_t^{\text{assimilation}} \) updates the subsystem parameter:

\begin{equation}
P(O_t \mid S_t) = \prod_{i=1}^{n} \mathcal{N}(O_t[i]; \mu_i, \sigma_i^2 / \sqrt{\kappa})
\end{equation}

where \( \mathcal{N}(x; \mu, \sigma^2) \) represents the normal distribution with mean \( \mu \) and variance \( \sigma^2 \). These values are derived from the interpolated dataset, with \( \kappa \) as the adjusted scale factor (Appendix \ref{appendix:scaling_factor}).
The updated digital state and quantity of interest complete the reward via the evaluation factor \( \phi_t^{\text{evaluation}} \). The reward function considers the control input, state transition, and observable data predictions to evaluate the performance of the DT-HA model.

For each control action \( u_t \), with a generalized probability \( P(u_t) = \frac{1}{n} \) (where \( n \) is the total number of control actions), the control reward \( R_{\text{control}} \) is as follows:

\begin{equation}
R_{\text{control}} = \frac{1}{k} \sum_{i=1}^{k} f(\psi_i^c)
\end{equation}

\noindent where \( k \) denotes the number of configurations, and \( f(\psi_i^c) \) is defined as follows:

\[
f(\psi_i^c) = 
\begin{cases} 
\left| \frac{1}{n} - \psi_i^c \right| & \text{if } \psi_i^c \neq \frac{1}{n} \\
-\epsilon & \text{if } \psi_i^c = 0 \\
\frac{1}{n} & \text{otherwise}
\end{cases}
\]

\noindent where, \( \epsilon \) is a small penalty for zero probability, and \( \frac{1}{n} \) represents the baseline probability. This function evaluates the deviation from the baseline probability, with penalties and adjustments ensuring effective control actions. The control reward \( R_{\text{control}} \) averages these values, measuring the effectiveness of the control action.

The state transition reward (\( R_{\text{state}} \)) for each step is calculated by comparing the derived probabilities to a baseline probability of 1.0:

\begin{equation}
R_{\text{state}} = \frac{1}{k} \sum_{i=1}^{k} f(\psi_i^s)
\end{equation}

\noindent where \( f(\psi_i^s) \) represents a piecewise function defined as follows:

\[
f(\psi_i^s) = 
\begin{cases} 
\left| 1.0 - \psi_i \right| & \text{if } \psi_i^s \neq 1.0 \\
-1 & \text{if } \psi_i^s = 0 \\
1.0 & \text{otherwise}
\end{cases}
\]

This function evaluates the deviation from the baseline probability. If \( \psi_i^s \) is not $1.0$, the absolute difference \( \left| 1.0 - \psi_i^s \right| \) is used and if \( \psi_i^s \) is $0$, a penalty of $-1$ is applied. Finally, if \( \psi_i^s \) equals the baseline, $1.0$ is added.

The observation prediction reward (\( R_{\text{obs}} \)) for each step is calculated by comparing the derived probabilities to a baseline probability of \( \frac{1}{n} \) as follows:

\begin{equation}
R_{\text{obs}} = \frac{1}{k} \sum_{i=1}^{k} f(\psi_i^o)
\end{equation}

\noindent where \( f(\psi_i^o) \) is defined as follows:

\[
f(\psi_i^o) = 
\begin{cases} 
\left| \frac{1}{n} - \psi_i^o \right| & \text{if } \psi_i^o \neq \frac{1}{n} \\
-0.1 & \text{if } \psi_i^o = 0 \\
\frac{1}{n} & \text{otherwise}
\end{cases}
\]

This function evaluates the deviation from the baseline probability. If \( \psi_i^o \) is not \( \frac{1}{n} \), the absolute difference \( \left| \frac{1}{n} - \psi_i^o \right| \) is used. If \( \psi_i^o \) is $0$, a penalty of $-0.1$ is applied, and if \( \psi_i^o \) equals the baseline, \( \frac{1}{n} \) is added.

This approach ensures that the reward calculations for control, state transition, and observation prediction are flexible and adaptable to various configurations.

\textbf{Evaluation}. During the operational phase, the DT assimilates subsystem data, estimates power generation parameters, and responds with appropriate control inputs. The prediction probabilities of the DT for \(D_t\), \(O_t\), \(U_t\), and \(R_t\) are considered.

\begin{figure}[!t]
  \centering
  \includegraphics[height=3.7cm, width=0.4\textwidth]{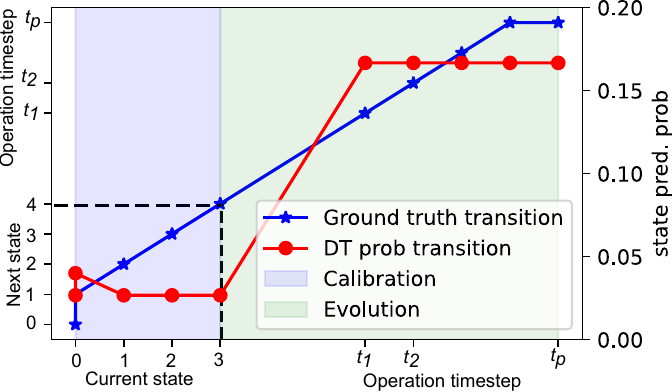}
  \caption{State transition evaluation}
  \label{state_eval}
\end{figure}

Fig. \ref{state_eval} illustrates the transition probabilities of the DT for digital states \(D_0\) to \(D_3\), representing the system cooling and venting, cold shutdown, hot shutdown and hot standby (Appendix \ref{appendix:state_transition_prob}). These states mark the calibration phase, after which energy production begins. Beyond state \(D_3\), the DT predicts and applies control inputs for energy production during timesteps \(t_1\) to \(t_p\).

The DT state prediction probabilities for \(D_t\) were evaluated using three configurations: \(P(D_t \mid D_{t-1})\), \(P(D_t \mid D_{t-1}, U_{t-1}, O_t)\), and \(P(D_t \mid D_{t-1}, D_t, O_t)\). The first configuration uses only previous states. The second includes control inputs and current observations, and the third combines past and current states with observable data for comprehensive predictions. As depicted in Fig. \ref{state_eval}, the DT prediction confidence increases as more data are assimilated. During calibration, probabilities stay below 0.05, indicating initial low knowledge. As operational data are assimilated, the confidence and accuracy of the DT improve, reaching beyond a probability of 0.15.

Next, the observation prediction probabilities of the DT are evaluated using five configurations: \(P(O_t \mid D_t)\), \(P(O_t \mid D_{t-1}, D_t)\), \(P(O_t \mid D_{t-1}, D_t, O_t)\), \(P(O_t \mid O_{t-1}, D_t)\), and \(P(O_{t+1} \mid D_{t+1}, D_t)\). These configurations assess predictions based on the current state, previous and current states, states with current observations, previous observations with the current state, and future observations based on state transitions, respectively. These configurations monitor the prediction accuracy of the DT and demonstrate the system's evolution with data assimilation (Appendix \ref{appendix:sobs_transition_prob}).

\begin{figure}[ht]
  \centering
  \begin{subfigure}{1\columnwidth}
    \includegraphics[height=3.5cm, width=\linewidth]{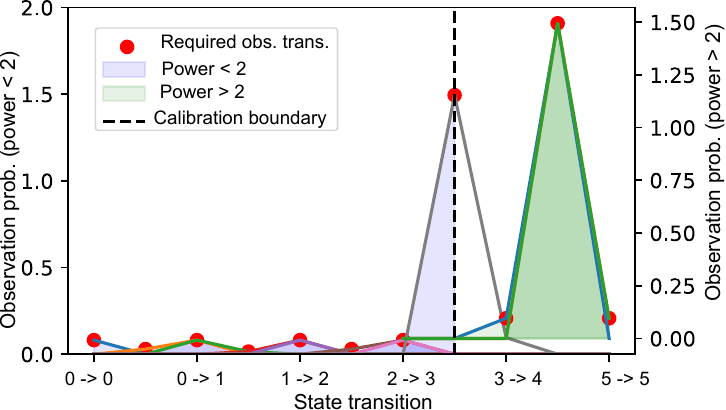}
    \caption{}
    \label{fig:obs_eval_1}
  \end{subfigure}
  \hfill
  \begin{subfigure}{1\columnwidth}
    \includegraphics[height=3.5cm,width=\linewidth]{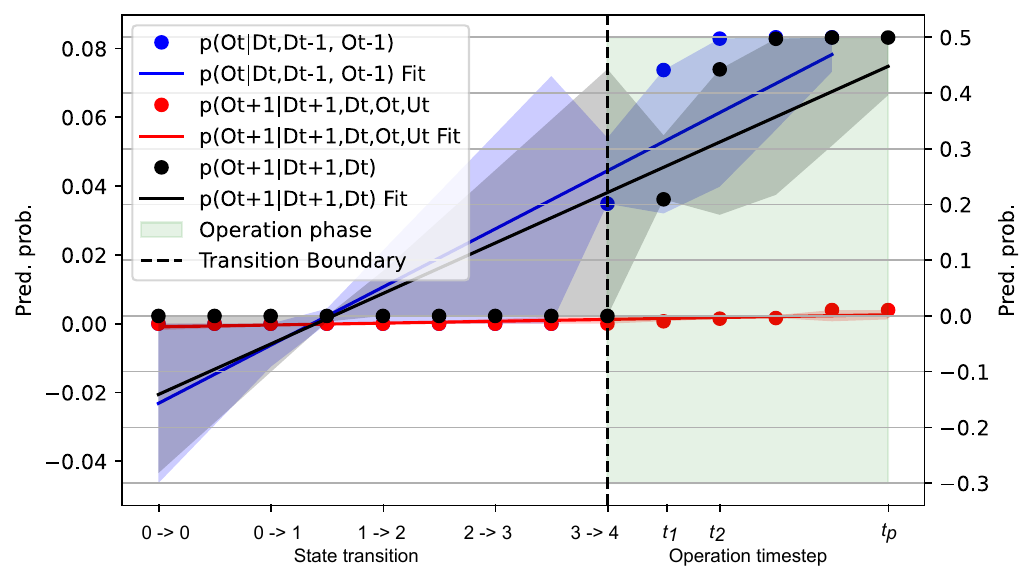} 
    \caption{}
    \label{fig:obs_eval_2}
  \end{subfigure}
  \caption{Observed data evaluation: (a) Evolution (b) Prediction}
  \label{fig:perf_obs_eval}
\end{figure}

During calibration, the digital state remaining in its current state versus transitioning to the next state is considered. As presented in Fig. \ref{fig:perf_obs_eval}(a), the model initially has low probabilities for transitioning due to the limited data at calibration. Despite this, the transition probability is higher than remaining in the current state. Beyond calibration, as more data are assimilated, the probabilities for transitioning to the next state increase, improving accuracy.

In the prediction evaluation in Fig. \ref{fig:perf_obs_eval}(b), the model prediction confidence improves with increased data assimilation. Scenarios \(P(O_t \mid D_t, D_{t+1}, O_{t+1})\) and \(P(O_{t+1} \mid D_{t+1}, D_t)\) exhibit higher probabilities and better accuracy. Conversely, \(P(O_{t+1} \mid D_{t+1}, D_t, O_t, U_t)\) maintains a low probability beyond 2\% operation, suggesting overfitting. Thus, except for \(P(O_{t+1} \mid D_{t+1}, D_t, O_t, U_t)\), prediction probabilities significantly improve beyond 2\% operation, indicating the enhanced predictive ability of DTs with more data assimilation, reflecting system evolution.

\begin{figure}[!t]
  \centering
  \begin{subfigure}{0.7\columnwidth}
    \includegraphics[height=3.5cm,width=\linewidth]{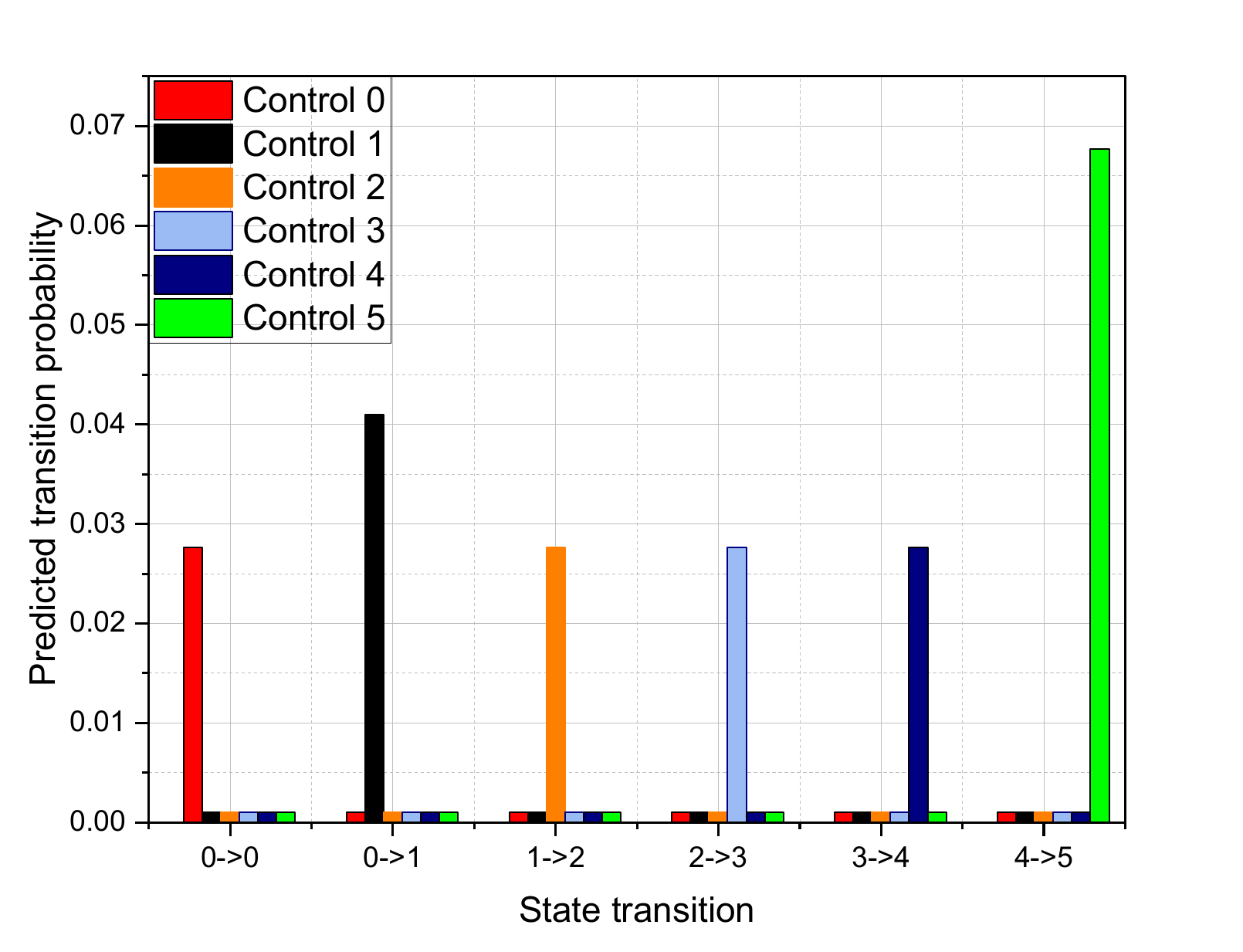}
    \caption{}
    \label{fig:cntr_eval_1}
  \end{subfigure}
  \hfill
  \begin{subfigure}{1\columnwidth}
    \includegraphics[height=3.5cm, width=\linewidth]{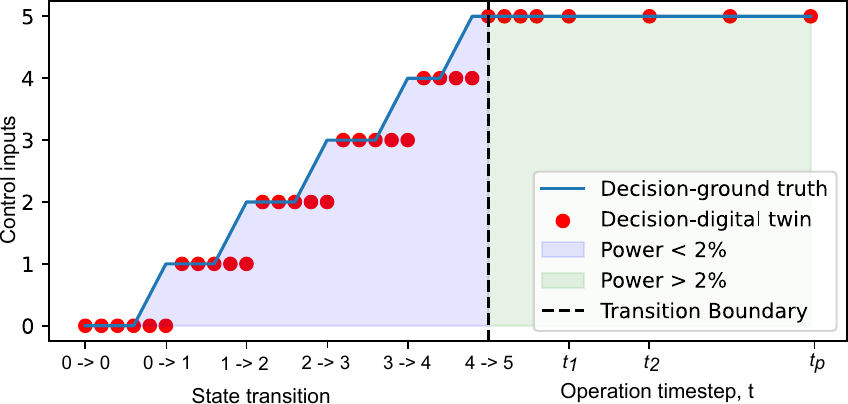} 
    \caption{}
    \label{fig:cntr_eval_2}
  \end{subfigure}
  \caption{Control input evaluation: (a) Calibration (b) Prediction}
  \label{fig:perf_cntr_eval}
\end{figure}

 Furthermore, we evaluate the control input calibration probabilities for each control input and the next state (Appendix \ref{appendix:control_input_transition_prob}). The DT intelligently determines the required control input for all transitions, as illustrated in Fig. \ref{fig:perf_cntr_eval}(a). The control input with the highest probability is activated, ensuring the correct action is triggered to drive the system to the next required state. For the control input prediction probabilities, we consider six configurations: \(P(u_t \mid S_{t-1}, S_t)\), \(P(u_t \mid S_t)\), \(P(u_t \mid S_{t-1}, S_t, O_t)\), \(P(u_t \mid S_t, O_t)\), \(P(u_t \mid S_{t-1}, S_t, O_t)\), and \(P(u_t \mid S_{t-1}, S_t, O_{t-1})\). These configurations assess the influence of the control actions based on the state transitions, next state predictions, states with observable data, and past observable data. Fig. \ref{fig:perf_cntr_eval}(b) presents the activated control inputs over state transitions during prediction. During the operation phase, the DT accurately issues the necessary control actions, maintaining a control input of 5 for energy prediction from \(t_1\) to \(t_p\).  

\begin{figure}[!t]
  \centering
  \includegraphics[height=3.7cm,width=0.45\textwidth]{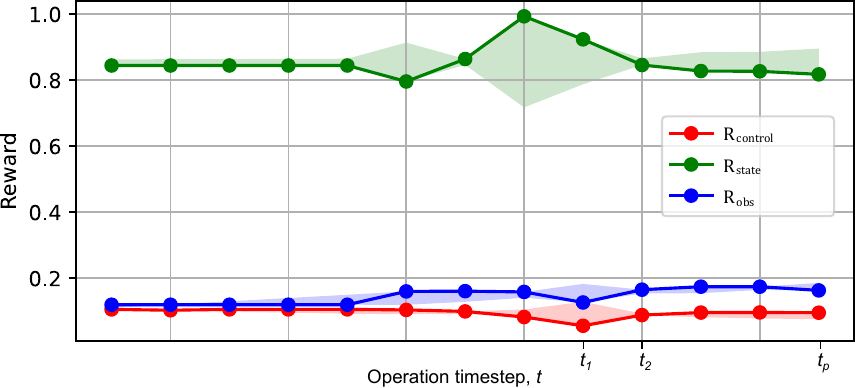}
  \caption{Reward evaluation}
  \label{fig:reward_eval}
\end{figure}

 Fig. \ref{fig:reward_eval} presents the reward evaluation. The control reward values (\(R_{\text{control}}\)) are low and steady, with a noticeable decrease at transition 9 before normalizing, indicating occasional low reward but generally moderate uncertainty in predicting control actions. Although the model displays potential for automation, human oversight remains necessary. The observation reward (\(R_{\text{obs}}\)) values are also moderate, with noticeable changes at transitions 6 and 9,  followed by normalization. This finding indicates a stable prediction accuracy for observable data with occasional uncertainty. The model reliably predicts general trends, which is crucial for operational monitoring. Minor variations suggest some variability but overall robust performance. The state reward values (\(R_{\text{state}}\)) are consistently high, peaking around transitions 7 to 9. This demonstrates excellent predictive accuracy for state transitions, making the model dependable for monitoring and planning.
 
 The DT model indicates a strong potential for state replication, with moderate success in control and observation predictions. Although the state predictions are highly reliable, the control and observation predictions require refinement for improved accuracy.

\subsection{Performance Evaluation of the Game-theoretic Offloading Framework}
This section presents the numerical results and analysis of the simulation to evaluate the performance of the proposed model. The C-MEC network is considered, where the subsystem is randomly distributed in $200 \, \text{m} \times 200 \, \text{m}$ area with $[4,12]$ subsystems, $[1,10]$ COIN nodes, and an ES. The large-scale fading from the $m$th subsystem to the $k$th AP is modeled as $g_m = 10^{\left(\frac{PL(d_{mk})}{10}\right)}$ with path loss $PL(d_{mk}) = -35.3 - 37.6 \log_{10}(d_{mk}) = -35.3 - 37.6$ \cite{van2022urllc}. The spectral density of the noise is set to $-174 \, \text{dBm/Hz}$ \cite{nasir2020resource}, and the bandwidth is set to $10 \, \text{MB}$. The URLLC decoding error probability is $\epsilon = 10^{-9}$. Table \ref{table:sim-param} provides additional parameters.

\begin{table}[ht]
    \centering
    \caption{Simulation Parameters}
    \label{table:sim-param}   
    \scriptsize
    \begin{tabular}{ll}
        \hline
        \textbf{Parameters} & \textbf{Value} \\
        \hline
        $I_m$  & [1, 10] MB\\
         $C_m$  , and  $T_m^{max}$ & [0.001, 0.1] GHz; 15 ms\\
        MEC capacity & 30 GHz \cite{pham2022partial} \\
        COIN node capacity & [1, 10] GHz \cite{lia2022in} \\
        Transmission power of subsystems &  0.5 W \\
        Unit gain of latency reduction & 2.5\\
        Offloading cost per workload & $0.1  \times 10\text{GHz}$\\
        Experience memory & $10000$\\
        Discount factor & 0.9\\
        \hline
    \end{tabular}
\end{table}

To verify the effectiveness of the proposed method, we evaluated the proposed approach against the following baselines:

\begin{itemize}
  \item \textit{The proposed scheme (DDQN-EPG)} employs the DDQN to predict the future optimal ORRA in a game-theoretic framework based on an EPG to maximize the user utility in a C-MEC network.
  
  \item \textit{The EPG with random ORRA (EPG-Rand)} strategy is based on randomly predicted future ORRA. This baseline offers insight into the overall future system performance when the DDQN is not applied.
  
  \item \textit{The MEC baseline} is the conventional MEC network with no COIN capabilities enabled, in which the subsystem can perform the task locally or offload it to the MEC. This baseline allows for a direct comparison between the proposed COIN approach and the standard MEC baseline, highlighting the performance improvement.
\end{itemize}

\begin{figure}[ht]
  \centering
  \begin{subfigure}{0.48\columnwidth}
    \includegraphics[width=\linewidth]{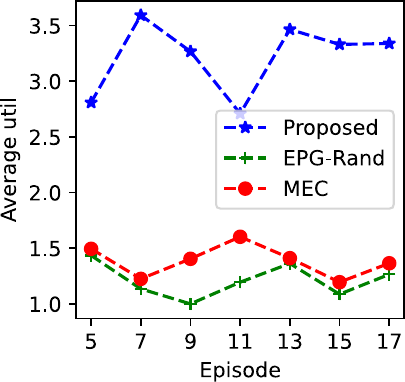}
    \caption{}
    \label{fig:subfig1a}
  \end{subfigure}
  \hfill
  \begin{subfigure}{0.48\columnwidth}
    \includegraphics[width=\linewidth]{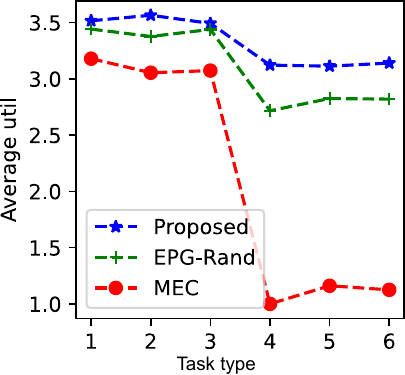} 
    \caption{}
    \label{fig:subfig1b}
  \end{subfigure}
  \caption{Performance evaluation based on experimental parameters on the system model (a) Episode. (b) Task input type.}
  \label{COIN-Ep_Task}
\end{figure}

\begin{figure}[ht]
  \centering
  \begin{subfigure}{0.48\columnwidth}
    \includegraphics[width=\linewidth]{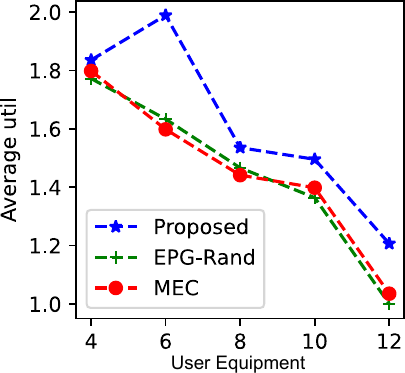}
    \caption{}
    \label{fig:subfig2a}
  \end{subfigure}
  \hfill
  \begin{subfigure}{0.48\columnwidth}
    \includegraphics[width=\linewidth]{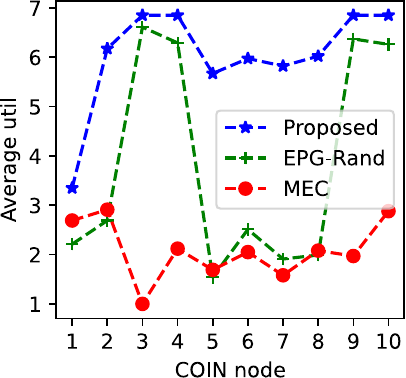} 
    \caption{}
    \label{fig:subfig2b}
  \end{subfigure}
  \caption{Performance evaluation based on user/in-network (COIN) node. (a) User equipment number. (b) COIN nodes.}
  \label{COIN-UE_COIN}
\end{figure}
\vspace{0.2em}
To ensure a fair performance comparison, we conducted a comprehensive analysis of various aspects. The evaluation involved comparing the average system utility across training episodes against the benchmark scenarios. Except for Episode 5, the proposed model consistently achieved the highest average system utility, as presented in Fig. \ref{COIN-Ep_Task}(a). In a few episodes, the proposed scheme attains 20\% utility over the baseline, demonstrating an effective offloading ratio and resource management via the DDQN.

Furthermore, the effectiveness of the proposed system model was evaluated by investigating the influence of data-intensive and computationally intensive computing types. For data-intensive tasks (Tasks 1 to 3), the input size ($I_m$ ) and required CPU cycles ($C_m$ ) of the tasks were uniformly and randomly generated from the ranges of [10-20] MB and [0.1-0.5] GB, respectively. In the computationally intensive task type, Im and Cm were uniformly and randomly generated from the ranges [1-5] MB and [1-2] GB, respectively. Considering the average system utility, the proposed model consistently outperformed the others, with an increase of 43.0\% to 87.9\% for data-intensive tasks (1–3) and 36.2\% to 87.7\% for computationally intensive tasks (4–6) compared to the second-best MEC model, as presented in Fig. \ref{COIN-Ep_Task}(b).

Next, we evaluated the performance of the proposed model by investigating the influence of varying the numbers of pieces of subsystem and COIN nodes. For subsystem numbers, the proposed method consistently demonstrates superior utility and effectiveness in Fig. \ref{COIN-UE_COIN}(a), underscoring a remarkable 47\% increment in utility compared to the baselines. Notably, an increase in subsystem beyond six resulted in an overall reduction in the average system utility. Per the COIN node numbers in Fig. \ref{COIN-UE_COIN}(b), the proposed model excels with a significant 64\% improvement over the baselines for between five and eight COIN nodes. Although the EPG-Rand significantly improves beyond eight COIN nodes, the proposed model maintains improved performance, underscoring the ability of the proposed approach to enhance the OPG algorithm, making it more efficient for increasing COIN-enabled nodes.

\section{Conclusion}
\label{sec: conclusion}
This paper investigated integrating DTs for HA (the DT-HA model) with PCO in a 6G-powered C-MEC environment. We proposed a decentralized game-theoretical approach to minimize the PCO latency by optimizing offloading decisions, ratios, and resource allocation. A PGM captured the intricacies of the DT-HA model, guiding HAs through observable data and control inputs. Offloading decisions were modeled as a strategic game, determining the subsystem utility under HA constraints. A DDQN refined the ORRA. The proposed approach effectively captures complex HAs while optimizing resource allocation, ensuring safe and efficient NPP operations. Future work will explore precise control actions from various NPP subsystems under practical operations, enhancing the robustness and applicability of the proposed method while addressing control complexity and security threats.

{\appendix
\subsection{State Transition Probabilities}
\label{appendix:state_transition_prob}

State transition probabilities define the likelihood of transitioning from one state to another represented in matrix form where each entry \( P(S_{t+1} = j \mid S_t = i) \) denotes the probability of moving from state \( i \) to  \( j \).

\[
S_t =
\left[
\begin{array}{ccccccccc}
0.4 & 0.6 & 0.0 & 0.0 & 0.0 & 0.0 & \cdots & 0.0 & 0.0 \\
0.0 & 0.4 & 0.6 & 0.0 & 0.0 & 0.0 & \cdots & 0.0 & 0.0 \\
0.0 & 0.0 & 0.4 & 0.6 & 0.0 & 0.0 & \cdots & 0.0 & 0.0 \\
0.0 & 0.0 & 0.0 & 0.4 & 0.6 & 0.0 & \cdots & 0.0 & 0.0 \\
0.0 & 0.0 & 0.0 & 0.0 & 0.4 & 0.6 & \cdots & 0.0 & 0.0 \\
0.0 & 0.0 & 0.0 & 0.0 & 0.0 & 0.4 & \cdots & 1.0 & 1.0 \\
\end{array}
\right]
\]

\subsection{Control Input}
\label{appendix:control_input_transition_prob}

The control input distribution \( U_t \) assigns equal probabilities to each possible control action. The control input distribution \( U_t \) is defined as follows:
\begin{equation}
\begin{split}
P(U_t = \text{u}_i) = \frac{1}{n}, \quad i \in \{0, 1, 2, \ldots, n-1\}, \\
n \in \mathbb{N} \text{ (total number of control actions)}. \nonumber
\end{split}
\end{equation}

\noindent Given the control actions, the conditional probability table for control input \(P(U_t \mid S_{t-1}, S_t)\) can be generalized as follows:

For \( S_{t-1} = (i, i) \) where \( i = 0, 1, 2, 3, 4, 5 \):
\[ 
P(U_t \mid (i, i), S_t) = 
\begin{cases} 
0.4 & \text{if } S_t = (i, i) \land U_t = u_i \\
0.6 & \text{if } S_t = (i, i+1) \land U_t = u_{i+1} \\
\end{cases}
\]

For \( S_{t-1} = (i, i+1) \) where \( i = 0, 1, 2, 3 \):
\[ 
P(U_t \mid (i, i+1), S_t) = 
\begin{cases} 
0.1 & \text{if } S_t = (i, i+1) \land U_t = u_{i+1} \\
0.4 & \text{if } S_t = (i+1, i+1) \land U_t = u_{i+1} \\
\end{cases}
\]

For \( S_{t-1} = (4, 5) \):
\[ 
P(U_t \mid (4, 5), S_t) = 
\begin{cases} 
0.4 & \text{if } S_t = (5, 5) \land U_t = u_5 \\
\end{cases}
\]

For \( S_{t-1} = (5, 5) \):
\[ 
P(U_t \mid (5, 5), S_t) = 
\begin{cases} 
1.0 & \text{if } S_t = (5, 5) \land U_t = u_5 \\
\end{cases}
\]
This representation captures the probabilities of control actions given the previous state \( S_{t-1} \) and current state \( S_t \). Any combination not listed has a probability of 0.

\subsection{Observable Data Transition Probability}
\label{appendix:sobs_transition_prob}
We let \( O_t \) represent the observable data at time \( t \). Each \( O_t \) consists of six key parameters: the PL, RCT, RCP, SGP, SGL, and RP. The probability distribution of the observable data \( O_t \) is represented as follows:

\[
P(O_t) = 
\begin{cases} 
\frac{1}{n_0} & \text{if } O_t \in \{O_0, O_1, O_2, O_3, O_4\} \\
\frac{f(O_t)}{\sum_{k>4} f(O_k)} & \text{if } O_t \in \{O_5, O_6, \ldots, O_p\}
\end{cases}
\]

\noindent where \( n_0 = 6 \). The first condition applies during calibration, and the second condition applies beyond state 4 in the operational phase based on design needs. Each \( O_t \) at time \( t \) includes parameters such as PL, RCT, RCP, SGP, SGL, and RP, with probabilities reflecting system specifications. For instance, the probability of observing \( [100, 60, 27, 1, 100, 0] \) is \(\frac{1}{n_0} \). Irrelevant observable data can be set to $0$.

\subsection{Scaling Factor}
\label{appendix:scaling_factor}

The scaling factor (\( \kappa \)) in NPP operations quantifies uncertainty at various operational stages. In the initial states (\( S_t = (0, 0) \) and \( S_t = (0, 1) \)), the scaling factor is higher (3.5 and 2.5), reflecting greater uncertainty. As the system transitions to the intermediate states (\( S_t = (1, 2) \) and \( S_t = (2, 3) \)), the scaling factor decreases (2.0 and 2.5), indicating increased stability. In more stable states (\( S_t = (3, 4) \)), the factor reduces to 1.0. At high power levels (\( S_t = (5, 5) \) to \( S_t = (5, 10) \)), the factor is 0, reflecting minimal uncertainty. This sequential adjustment captures the dynamic uncertainty across different operational stages of the NPP.

\subsection{Dynamic Estimation}
\label{append: Dynamic_Estimation}
\subsubsection{Planning and Optimal Control} 
\label{append: plan_control}
At each time step, the DT of the NPP selects optimal control inputs in response to evolving operational conditions. For instance, maintaining the optimal reactor temperature and safe SGP levels can be formulated as a planning problem. In another instance, the DT can ensure safe operation by adjusting the control rod positions dynamically to balance the power output and fuel usage of the reactor.
The segment from the current time step \( t_c \) to a future time step \( t_p \) can be viewed as a partially observable MDP. The goal is to select control inputs \( u_{t_c}, \ldots, u_{t_p} \) that maximize the expected future reward. The control policy \( \pi \)  maps from the current belief over the entire history to a control action, as follows \cite{kapteyn2021}:

\begin{equation}
u_t = \pi \left( p(D_0, \ldots, D_t, Q_0, \ldots, Q_t \mid o_0, \ldots, o_t, u_0, \ldots, u_{t-1}) \right).
\end{equation}

The objective is to determine a policy that maximizes the expected accumulated reward:

\begin{equation}
\pi^* = \arg \max_\pi \sum_{t=t_c+1}^{t_p} \gamma^{(t-t_c-1)} E[R_t],
\end{equation}

\noindent where \( \gamma \in [0, 1] \) denotes a discount factor.If state estimates are accurate, the partially observable MDP is approximated as a fully observable MDP. The expected policy is

\begin{equation}
u_t = \tilde{\pi} (\hat{d}, \hat{q}),
\end{equation}

\noindent  where \(\hat{d}\) and \(\hat{q}\) represent the best estimates of the current state and quantities of interest. This problem is solved off-line using the value iteration algorithm. The reward function is

\begin{equation}
R_t(u_t, q_t) = R_{\text{state}}(q_t) + R_{\text{control}}(u_t) + R_{\text{obs}}(o_t),
\end{equation}

\noindent with a discount factor \( \gamma = 0.6 \).

\subsubsection{Extension to Prediction} 
\label{append: Extending_predic}
The NPP DT formulation is extended to include prediction over future time steps. The prediction regime spans from the current time step, \( t_c \), to a chosen prediction time step,  \( t_p \), in the PGM. The target belief state is expanded to predict the digital state, quantities of interest, and reward variables up to the prediction horizon, \( t_p \), as follows \cite{kapteyn2021}:\

\begin{equation}
\begin{split}
p(D_0, \ldots, D_{t_p}, Q_0, \ldots, Q_{t_p}, R_0, \ldots, R_{t_p}, \\
U_{t_c+1}, \ldots, U_{t_p} \mid o_0, \ldots, o_{t_c}, u_0, \ldots, u_{t_c}) \\
\propto \prod_{t=0}^{t_p} \left[ \phi^{\text{dynamics}}_t \phi^{\text{QoI}}_t \phi^{\text{evaluation}}_t \right] \\
\prod_{t=0}^{t_c} \phi^{\text{assimilation}}_t \\
\prod_{t=t_c+1}^{t_p} \phi^{\text{control}}_t
\end{split}
\end{equation}

\noindent where

\[
\phi^{\text{control}}_t = p(U_t \mid D_t, Q_t).
\]

The additional required term is the control factor, \(\phi^{\text{control}}_t\), defined according to the control policy as

\[
p(u_t \mid d_t, q_t) = \begin{cases} 
1 & \text{if } \tilde{\pi}(d_t, q_t) = u_t, \\
0 & \text{otherwise}.
\end{cases}
\]

This expression represents the probability of selecting a control action  \( u_t \) given the state \( d_t \) and quantities of interest \( q_t \). The policy \( \tilde{\pi}(d_t, q_t) \)  determines the optimal control action. If the selected control action \( u_t \) matches the policy \( \tilde{\pi}(d_t, q_t) \), the probability \( p(u_t \mid d_t, q_t) \) is 1; otherwise, it is 0. This approach ensures that only the optimal control action has a nonzero probability.
In the prediction regime for \( t = t_c + 1, \ldots, t_p \), the data assimilation factor is omitted and not conditioned on \( O_t = o_t \) because future data \( o_t \) are not yet observed. These adjustments and additional control factor enable seamless prediction and monitoring in a single pass of the sum-product algorithm.

\subsection{Proof of Remark 1}
\label{appendix: NE_prove}
The PCO computation offloading game is formulated under DT-HA constraints for the decision-making process of subsystems regarding whether to offload tasks. Based on the definition of EPG in \cite{pham2022partial}, the game  \( G \) should satisfy the following condition. 
Subsystem \( m \) is considered, which alters its strategy to demonstrate that this game is an EPG and achieves NE as follows:

\begin{equation}
\begin{gathered}
\phi(s_m,s_{-m}) - \phi(s_m',s_{-m}) = U_m(s_m,s_{-m}) - U_m(s_m',s_{-m}), \\ \forall m \in M, \forall s_m, s_m' \in S_m.
\end{gathered}
\end{equation}

For ease of proof, the expression \(\phi(s_m,s_{-m})\) is given as follows:

\begin{align}
\phi(s_m,s_{-m}) &= s_{m0} \sum_{m \in M} R_{m0} \nonumber \\
&\quad + (1 - s_{m0}) \left( \sum_{j \in K} s_{mj} R_{mj} + \sum_{m' \neq m} R_{m'0} \right), \label{eq:append_EPG_potential_func}
\end{align}

The relationship between \(\phi(s_m,s_{-m}) - \phi(s_m',s_{-m})\) and \(U_m(s_m,s_{-m}) - U_m(s_m',s_{-m})\) is discussed in two cases as follows.

\textit{Case 1:} Transitioning from MEC (i.e., \(s_{m0} = 1\)) offloading to CN nodes (i.e., \(s_{mj} = 1\)). Using Eq.\ref{eq:append_EPG_potential_func}, we deduce that 

\begin{align}
\phi(s_{m},s_{-m}) - \phi(s_{mj},s_{-m}) &= \sum_{m \in M} R_{m0} \nonumber \\
&\quad - \left( R_{mj} + \sum_{m' \neq m} R_{m'0} \right) \nonumber \\
&= R_{m0} - R_{m'0} \nonumber \\
&= U_m(s_{m0},s_{-m}) - U_m(s_{mj},s_{-m}), \label{eq:append_EPG_Case1}
\end{align}

Hence, (\ref{eq:append_EPG_Case1}) is established in this case.

\textit{Case 2:} Altering strategies only in the CN environment (i.e., from CN \( j \) to another CN \( j' \)), we have:

\begin{align}
\phi(s_{mj},s_{-m}) - \phi(s_{m j'},s_{-m}) &= R_{mj} - R_{m j'} \nonumber \\
&= U_m(s_{m0},s_{-m}) - U_m(s_{mj},s_{-m}), \label{eq:append_EPG_Case2}
\end{align}

Thus, (\ref{eq:append_EPG_Case2})  is established in this case.

Equation (\ref{eq:append_EPG_potential_func}) is established in any case from the above derivations. Hence, game \( G \) is an EPG and attains NE in a finite number of iterations.
}

\bibliographystyle{IEEEtran}
\bibliography{main}


\begin{IEEEbiography}[{\includegraphics[width=1in,height=1.25in,clip,keepaspectratio]{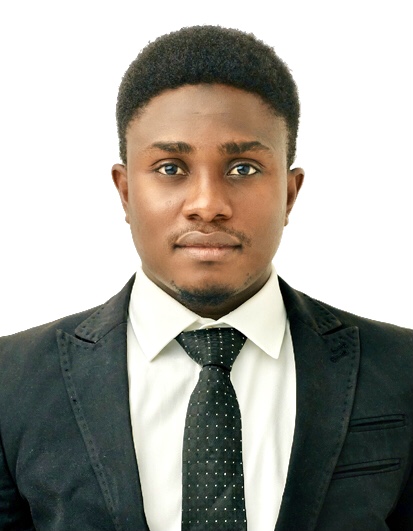}}]{Ibrahim Aliyu} (Member, IEEE) received a PhD in computer science and engineering from Chonnam National University in Gwangju, South Korea, in 2022. He also holds BEng (2014) and MEng (2018) degrees in computer engineering from the Federal University of Technology in Minna, Nigeria. He is currently a postdoctoral researcher with the Hyper Intelligence Media Network Platform Lab in the Department of Intelligent Electronics and Computer System Engineering at Chonnam National University. His research focuses on distributed computing for massive metaverse deployment. His other research interests include federated learning, data privacy, network security, and artificial intelligence for autonomous networks.
Use 
\end{IEEEbiography}

\begin{IEEEbiography}[{\includegraphics[width=1in,height=1.25in,clip,keepaspectratio]{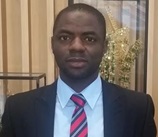}}]{Awwal M. Arigi}
 holds a PhD in nuclear system engineering, specializing in human and organizational factors. He is a research scientist at the Institute for Energy Technology, Norway. Previously, he worked as an assistant professor in the Nuclear Engineering department at Chosun University in South Korea, where he taught Probabilistic Risk Analysis and Human Factors Engineering in Nuclear Power Plants. He also spent over three years as a research assistant and postdoctoral researcher at the Human Engineering and Risk Analysis Laboratory. Dr. Arigi has participated in several industry-related projects on multiunit risk, human reliability analysis, and probabilistic safety analysis. His current research focuses on human-automation, digital twins, and small modular reactor operations, with several related publications.
\end{IEEEbiography}

\begin{IEEEbiography}[{\includegraphics[width=1in,height=1.25in,clip,keepaspectratio]{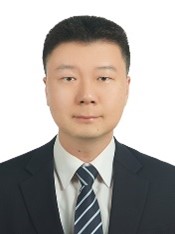}}]{Tai-Won Um}
received a BS degree in electronic and electrical engineering from Hongik University in Seoul, South Korea, in 1999 and MS and PhD degrees from the Korea Advanced Institute of Science and Technology (KAIST) in Daejeon, South Korea, in 2000 and 2006, respectively. From 2006 to 2017, he was a principal researcher with the Electronics and Telecommunications Research Institute (ETRI), a leading government institute on information and communications technology in South Korea. He is currently an associate professor at Chonnam National University in Gwangju, Korea. He has been actively participating in standardization meetings, including ITU-T SG20 (Internet of Things, smart cities, and communities).
\end{IEEEbiography}

\begin{IEEEbiography}[{\includegraphics[width=1in,height=1.25in,clip,keepaspectratio]{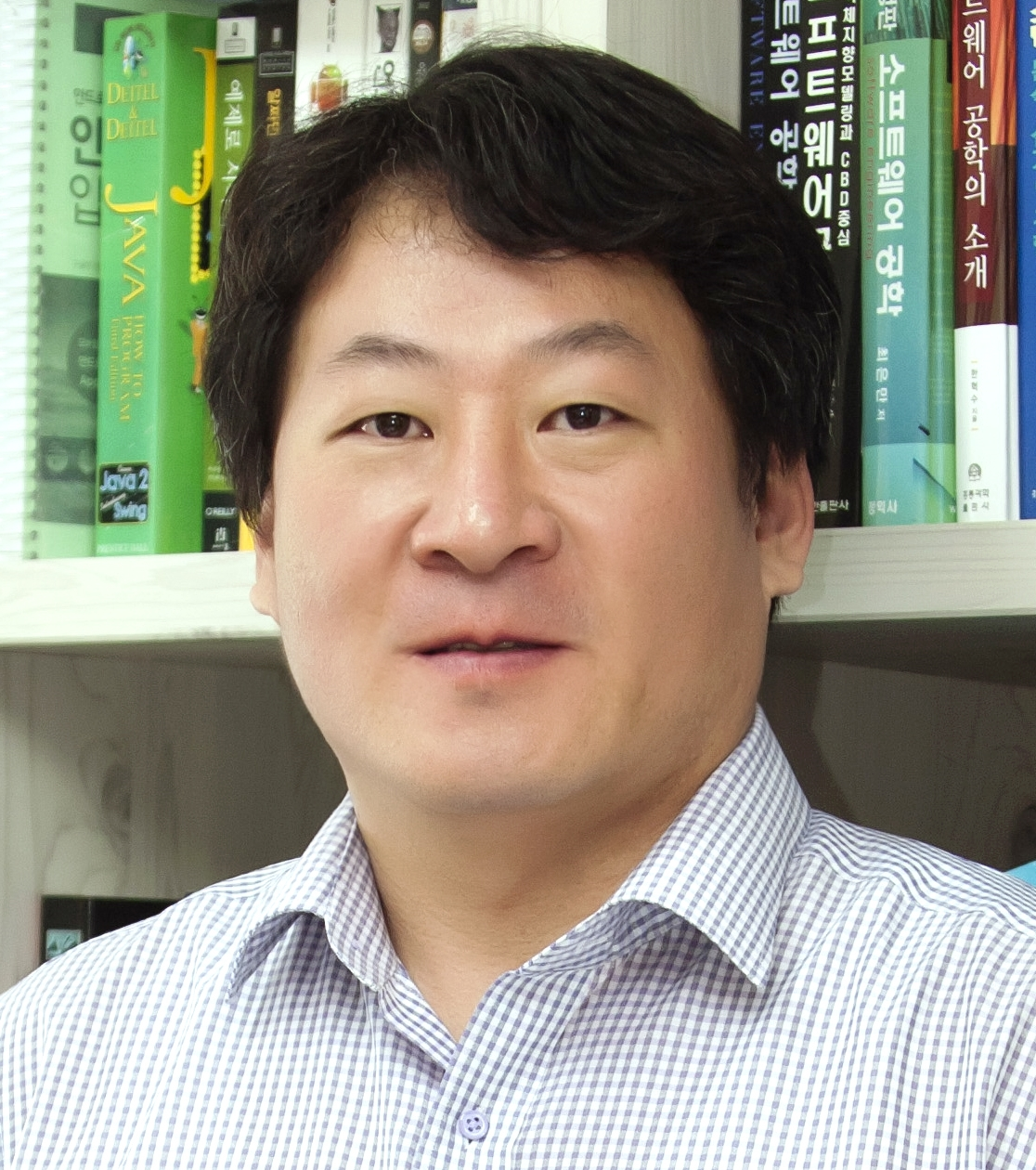}}]{Jinsul Kim}
(Member, IEEE) received a BS degree in computer science from the University of Utah at Salt Lake City in Utah, USA, in 1998 and MS (2004) and PhD (2008) degrees in digital media engineering from the Korea Advanced Institute of Science and Technology (KAIST) in Daejeon, South Korea. Previously, he worked as a researcher at the Broadcasting/Telecommunications Convergence Research Division, Electronics and Telecommunications Research Institute (ETRI) in Daejeon, Korea, from 2004 to 2009 and was a professor at Korea Nazarene University in Cheonan, Korea, from 2009 to 2011. He is a professor at Chonnam National University, Gwangju, Korea. He is a member of the Korean national delegation for ITU-T SG13 international standardization. He has participated in various national research projects and domestic and international standardization activities. He is a co-research director of the Artificial Intelligence Innovation Hub Research and Development Project hosted by Korea University and is the director of the G5-AICT research center.

\end{IEEEbiography}

\vfill

\end{document}